\documentclass[aps,pra,superscriptaddress,showpacs]{revtex4}
\usepackage[colorlinks=true,citecolor=blue]{hyperref}
\usepackage{graphicx}
\usepackage{epstopdf,epsfig}
\usepackage{amsmath}
\usepackage{amsfonts}
\usepackage{epstopdf,epsfig}
\usepackage{times}
\usepackage{color}
\usepackage[normalem]{ulem}
\usepackage{braket}
\usepackage{xcolor}

\usepackage{latexsym}
\usepackage{mathrsfs}
\usepackage{dsfont}
\usepackage{mathtools}
\usepackage{empheq}
\usepackage{mathbbol}
\usepackage{bbm}
\usepackage{siunitx}
\usepackage{calc}
\usepackage{comment}

\newcommand{\beq}{\begin{equation}}
\newcommand{\eeq}{\end{equation}}
\newcommand{\bea}{\begin{eqnarray}}
\newcommand{\eea}{\end{eqnarray}}

\newcommand{\m}{\mathrm}

\newcommand{\de}{\mathrm{d}}	

\newcommand{\signum}{\mathop{\mathrm{sgn}}}

\usepackage{hyperref}

\graphicspath{{pic}}

\begin{document}

\title{Effective Theory and Breakdown of Conformal Symmetry in a Long-Range Quantum Chain}

\author{L. Lepori}
\email[referent for correspondence at: ]{llepori81@gmail.com}
\thanks{ \href{mailto:luca.lepori@pd.infn.it}{luca.lepori@pd.infn.it}}
\affiliation{Dipartimento di Fisica e Astronomia, 
Universit\`a di Padova, Via Marzolo 8, I-35131 Padova, Italy}
\affiliation{icFRC, IPCMS (UMR 7504) and ISIS (UMR 7006), Universit\'{e} de Strasbourg and CNRS, Strasbourg, France}

\author{D. Vodola}
\affiliation{icFRC, IPCMS (UMR 7504) and ISIS (UMR 7006), Universit\'{e} de Strasbourg and CNRS, Strasbourg, France}

\author{G. Pupillo}
\affiliation{icFRC, IPCMS (UMR 7504) and ISIS (UMR 7006), Universit\'{e} de Strasbourg and CNRS, Strasbourg, France}

\author{G. Gori}
\affiliation{CNR-IOM DEMOCRITOS Simulation Center, Via Bonomea 265, I-34136 Trieste, Italy}

\author{A. Trombettoni}
\affiliation{CNR-IOM DEMOCRITOS Simulation Center, Via Bonomea 265, I-34136 Trieste, Italy}
\affiliation{SISSA and INFN, Sezione di Trieste, Via Bonomea 265, I-34136 Trieste, Italy}

\begin{abstract}
We deal with the problem of studying the symmetries and the effective theories of long-range models
around their critical points. A prominent issue is to determine whether they possess (or not) 
conformal symmetry (CS) at criticality and how the presence of CS depends on the range of the interactions. 
To have a model, both simple to treat and interesting, where to investigate these questions, we 
focus on the Kitaev chain with long-range pairings decaying with distance as power-law with exponent $\alpha$. 
This is a quadratic solvable model, yet displaying non-trivial quantum phase transitions. 
Two critical lines are found, occurring respectively at a positive and a negative chemical potential. 
Focusing first on the critical line at positive chemical potential,
by means of a renormalization group approach we derive its effective theory close to criticality. 
Our main result is that the effective action 
is the sum of two terms: a Dirac action $S_{\mathrm{D}}$, found in the short-range Ising universality class, 
and an ``anomalous'' CS breaking term $S_{\mathrm{AN}}$. While  $S_{\mathrm{D}}$ originates from 
low-energy excitations in the spectrum, $S_{\mathrm{AN}}$ originates from 
the higher energy modes where singularities develop, due to the long-range nature of the model. 
At criticality $S_{\mathrm{AN}}$ flows to zero for $\alpha > 2$, while for $\alpha < 2$ it dominates and 
determines the breakdown of the CS. 
Out of criticality $S_{\mathrm{AN}}$ breaks, in the considered approximation, 
the effective Lorentz invariance (ELI) for every finite $\alpha$. 
As $\alpha$ increases such ELI breakdown becomes less and less pronounced and 
in the short-range limit $\alpha \to \infty$ the ELI is restored. 
In order to test the validity of the determined effective theory, 
we compared the two-fermion static correlation functions and the von Neumann entropy 
obtained from them with the ones calculated on the lattice, finding agreement. 
These results explain two observed features characteristic of long-range models, the hybrid decay of static 
correlation functions within gapped phases and the area-law violation for the von Neumann entropy. 
The proposed scenario is expected to hold in other long-range models displaying 
quasiparticle excitations in ballistic regime. 
From the effective theory one can also see 
that new phases emerge for $\alpha<1$. 
Finally we show that
at every finite $\alpha$ the critical exponents, defined as for the short-range ($\alpha \to \infty$) model, 
are not altered. This also shows that the long-range paired Kitaev chain 
provides an example of a long-range model in which the value of $\alpha$ where the CS is  
broken does not coincide with the value at which the critical exponents start to differ 
from the ones of the corresponding short-range model. 
At variance, for the second critical line, having negative chemical potential,  
only $S_{\mathrm{AN}}$ ($S_{\mathrm{D}}$) is present for $1<\alpha<2$ (for $\alpha>2$). Close to this 
line, where the minimum of the spectrum coincides with the momentum where singularities develop, 
the critical exponents change where CS is broken. 
\end{abstract}

\maketitle

\section{Introduction} 
\label{intro}

The study of classical and quantum long-range systems, both at and out 
of equilibrium, is a very active field of research \cite{libro}. 
One of the main reasons for this growing interest
is that these systems have been predicted to
exhibit new phases with peculiar properties, including 
the presence of correlation functions with both exponential 
and algebraic decay, even in the presence of a mass gap 
\cite{cirac05,tagliais,nostro,paperdouble}, the 
violation of the lattice locality 
\cite{hast,hauke2013,eisert2014,metivier2014,noinf,damanik2014,nbound,storch2015,carleo,kastner2015ent,kuwahara2015} 
and of the area-law for the von Neumann entropy \cite{tagliais,nostro}, nonlinear growth of the latter quantity after a quench 
\cite{growth}, peculiar constraints on  
thermalization \cite{santos2015}, . 

Recently developed techniques in atomic, molecular and optical systems 
(such as Rydberg atoms, polar molecules, magnetic or electric dipoles, multimode cavities and trapped ions) 
provide an experimental playground 
to investigate the properties of phases and phase transitions for long-range models 
\cite{Childress2006,Balasubramanian2009,Weber2010,tech0,tech1,Schauss2012,Aikawa2012,Lu2012,Firstenberg2013,Yan2013,Dolde2013,Islam2013,exp1,exp2,tech5} 
and motivated an intense theoretical activity \cite{Saffman2010,tech2,tech3,exp0,tech4,maghrebi2015tris}. 
In particular, Ising-type spin chains with tunable long-range interactions can be now realized using neutral 
atoms coupled to photonic modes of a cavity or with trapped ions 
coupled to motional degrees of freedom. In this latter case, 
the resulting interactions decay algebraically with the distance $r$, 
with an adjustable exponent usually in the range $\alpha \lesssim 3$.

A crucial issue in the investigation of many-body systems is the characterization and the study of the symmetries 
of their critical points. For classical short-range 
models in two and three dimensions the scale invariance hosted at the critical 
points where second order phase transitions arise is expected to be promoted to the larger \emph{conformal symmetry} 
(CS), also including translations, Euclidean rotations, 
and special conformal transformations combining translations with spatial inversions \cite{dif,muss}.
This symmetry, conjectured long ago \cite{polyakov_1970}, 
has in two dimensions far-reaching consequences \cite{dif,muss}, 
as it fixes completely the universality class. 
Numerical studies both in two and three dimensions gave a clear evidence of the presence of CS for short-range models \cite{cardy_1994,langlands_1994,deng_2002,gori_2015,penedones_2015}. 

A related, natural question for long-range models is then 
whether their critical points possess (or not) 
CS and how its presence is eventually related to the range 
of the interactions. Referring to couplings decaying with the distance 
$r$ as $1/r^\alpha$, since for $\alpha \to \infty$ the short-range models and their CS at criticality are recovered, {a central issue is up} to what values of $\alpha$ (and how) CS persists.
This issue is related to another typical question arising in the study of long-range systems, where an important 
information is the determination of the upper limit value of $\alpha$ (say 
$\alpha^{\ast}$) such that {above it} the critical exponents of the short-range model are 
retrieved \cite{sak_1973,libro}.
For example, classical spin systems with long-range couplings having power-law exponent $\alpha \equiv d+\sigma$ display, for $\sigma$
greater than a critical value $\sigma^*$, the same critical
exponents of the corresponding short-range models ($\sigma \to \infty$), while for $d/2 <
\sigma \leq \sigma^*$, they exhibit peculiar long-range critical exponents \cite{libro,sak_1973}. \\
Finally, for classical long-range $O(N)$ models with continuous symmetry one may have a finite critical temperature also in one spatial dimension \cite{spohn1999}, this being not in contradiction with the Mermin-Wagner theorem \cite{MW,lebellac}, which is valid for short-rang interactions (see e.g. the discussion in \cite{defenu2014}). As a further example, the classical long-range Ising model in $d = 1$ \cite{dyson1969,thouless1969,anderson1970} has $\sigma^* = 1$ and
exactly at $\sigma= \sigma^*$, a phase transition of the Berezinskii-Kosterlitz-Thouless universality class occurs \cite{cardy1981,frolich1982,lui2001,lui1997}.

To deal with the interesting problems outlined above for interacting long-range quantum $1d$ chains or long-range classical 
$2d$ systems, namely the breakdown of the conformal symmetry at criticality and the behavior of the critical exponents, one has to eventually resort to computationally expensive numerical simulations. It is then clear that a qualitative understanding based 
on a simple, exactly solvable and possibly non-trivial model would be highly 
desirable.

The presence of CS at the critical point implies the Lorentz invariance of the theory. 
Typically, in short-range interacting models the resulting effective theory displays an 
\emph{effective Lorentz invariance} (ELI) also near 
criticality, as it is seen for the quantum Ising chain in a transverse field \cite{muss}. Moreover, 
general perturbations of the critical points of short-range models allow for 
Lorentz invariance also in massive regimes, where CS is broken \cite{muss}. 
It is then a natural question if and to what extent the ELI and associated locality survive, 
both at and out of criticality, in long-range critical systems and, when they are present, 
how they affect the physical observables, as correlation functions.

To shed light on the issues described above, 
we decided to consider a long-range quadratic fermionic model in one quantum dimension 
exhibiting non-trivial quantum phase transitions, the 
Kitaev chain with long-range pairings power-law decaying with distance, recently introduced in \cite{nostro}. 
In particular we derive, via a renormalization group (RG) approach, an effective continuous theory 
for this model at and close to its critical points.
 
The Kitaev with long-range pairing (and hopping) is already experimentally
realizable for $\alpha = 1$ for a particular type of  
helical Shiba chain \cite{Pientka2013,Pientka2014} (while for example for  
the long-range Ising model one can engineer $\alpha$ such that $\alpha \lesssim 3$ \cite{tech1,exp1,exp2}). 
Nevertheless, the long-range Kitaev chain provides an ideal playground to investigate the 
symmetries of the critical points of long-range systems, since: 
\begin{itemize}
\item It is a quadratic fermionic model, then exactly solvable. 
\item For $\alpha \to \infty$,  where the short-range Kitaev chain \cite{kitaev} is recovered, it  
is mapped via Jordan-Wigner transformations to the short-range Ising model in a 
transverse field \cite{libro_cha}  (a discussion of the comparison 
between the phases of the long-range Kitaev and long-range Ising chains can be found 
in \cite{paperdouble}). 
\item It displays non-trivial (non mean-field) 
quantum phase transitions at $\mu = \pm 1$, where 
$\mu$ is the chemical potential appearing in the Hamiltonian~\eqref{Ham}. 
\item As shown in \cite{nostro,paperdouble}, it exhibits, in the presence of vanishing mass gap, a 
linear spectrum around the zero-energy points, which is the  case  
considered in standard textbooks \cite{dif,muss}.
\item It displays a hybrid decay behavior (exponential at short distances and power-law at longer ones, 
see Fig.~\ref{fig:intro}) for the static correlation functions in gapped regimes \cite{nostro}. 
This hybrid behavior has also been observed e.g. in the Ising model with long-range interactions 
\cite{cirac05,paperdouble} and it seems to be characteristic of long-range systems.
\end{itemize}

In Ref.~\cite{paperdouble} the origin of the hybrid decay behavior mentioned above was identified 
analytically as the result of the competition between two different sets of modes with well-separated energies.
The short-distance exponential decay originates from the eigenmodes with energy near the minimum of the 
energy spectrum in the center of the Brillouin zone (as in short-range models), while, surprisingly, 
the long-distance algebraic decay originates from {\it high-energy} eigenmodes. 
A similar picture emerged for the long-range Ising model 
on the basis of a spin wave analysis \cite{cirac05}. 
This fact suggests that the double contribution to the correlation functions from two different sets of modes 
can be not limited to free or weakly interacting long-range models. 
In the long-range paired Kitaev chain the modes at the edges of the Brillouin zone 
have been found responsible also for an anomalous scaling for the ground state energy at criticality 
\cite{nostro,paperdouble}. For this reason, in this paper we put forward and develop 
an approximate renormalization group (RG) approach in which we explicitly take into account the contribution of the 
modes at the edges of the Brillouin zone during the decimation procedure.

Focusing mostly on the critical line $\mu = 1$, the main results that we obtained are:

\begin{figure*}
\includegraphics[width=0.3\textwidth]{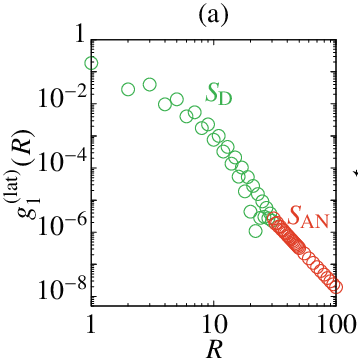}
\caption{Plot of the static two fermions correlation function 
$g_1^{\text{(lat)}}(R)$ defined in Eq.~\eqref{formcorr},  showing the hybrid exponential and power-law decay. 
The first part on the left is described close to criticality by the Dirac action $S_{\mathrm{D}}$, 
while the algebraic tail is associated to CS breaking term $S_{\mathrm{AN}}$ 
[see Section \ref{latcorr}].} 
\label{fig:intro}
\end{figure*}

\begin{itemize}

\item Section \ref{ET}: The resulting effective action is found to be the sum of two terms: 
a Dirac action $S_{\mathrm{D}}$ coming from contributions near the minimum of the energy spectrum, 
as in the short-range limit $\alpha \to \infty$, plus an anomalous CS breaking term $S_{\mathrm{AN}}$ 
originating from the higher-energy contributions at the edges of the Brillouin zone, 
where singularities develop (in the $[\alpha]$-th derivative of the spectrum, $[\alpha]$ labeling the 
integer part of $\alpha$) due to the long-range nature of the model. 
This is one of the central results of the paper. 
Although the determination of effective theories for both classical and quantum 
long-range models has been the subject 
of a perduring interest (see e.g. Refs.~\cite{laflorencie,defenu,Maghrebi2015} and references therein), 
in our opinion the present example 
is particularly instructive, since the precise origin and consequence of the anomalous CS breaking term 
$S_{\mathrm{AN}}$ is directly identified and derived from the microscopics. 
Notably the appearance of the CS breaking action, due to high-energy modes, 
implies a different hierarchy for the quasiparticles weights along the RG,
compared to the one generally assumed for short-range systems \cite{huang}. 

\item Sections \ref{ET} and \ref{csbreak}: At criticality, for $\alpha > 2$ 
RG makes the 
CS breaking term $S_{\mathrm{AN}}$ flow to zero, while for $\alpha<2$ it dominates over 
$S_{\mathrm{D}}$. This change of behavior is at the origin of the breakdown of CS, 
since $S_{\mathrm{AN}}$ is not conformal invariant.

\item Section \ref{ET}: The obtained effective theory allows to identify two new phases for $\alpha < 1$. 
Close to criticality these phases display an additional emergent
approximate symmetry under suitable anisotropic scale transformations.

\item Section \ref{latcorr}: Out of criticality $S_{\mathrm{AN}}$ and $S_{\mathrm{D}}$ do not decouple 
completely and they co-act to determine the physical quantities. For instance they 
are at the origin of the hybrid (exponential plus power-law) decay of the static two fermions correlation functions out of criticality. This coupling between the two terms $S_{\mathrm{D}}$ and $S_{\mathrm{AN}}$ and the validity 
of the effective theory are probed against the lattice results. 
The asymptotic decay exponents of the lattice correlation functions 
are compared to the ones obtained from the effective theory, finding perfect agreement.
Notably our study also unveils the central role played by RG subleading terms in $S_{\mathrm{AN}}$, 
affecting even qualitatively the correlation functions in some regimes. 
In our knowledge this fact has no counterpart in short-range systems.

\item Section \ref{lor_br}: The hybrid decay of the two fermions static correlation functions out of criticality 
points to the breakdown of the Lorentz invariance for the effective theory. 
Indeed we infer that out of criticality the ELI, exact in the limit $\alpha \to \infty$ 
(where $S_{\mathrm{D}}$ just acquires a mass term \cite{dif,muss}), is broken at every finite value 
of $\alpha$. Conversely, exactly at criticality the Lorentz group, belonging to the conformal group, 
is broken only below $\alpha = 2$.

\item Section \ref{VNE}: By means of the effective theory we computed the von Neumann entropy 
$V(\ell)$ ($\ell$ being the size of a part of the bi-parted chain). Exactly at criticality $V(\ell)$ 
follows the standard scaling law $V(\ell)= a + 
\frac{c^{\mathrm{(VNE)}}}{3} \log(\ell)$ valid for short-range critical systems 
\cite{wilczek,calabrese}. However, while above $\alpha = 1$ it is found $c^{\mathrm{(VNE)}} = \frac{1}{2}$, as 
for the short-range Kitaev (Ising) chain, below this threshold it holds $c^{\mathrm{(VNE)}}= 1$. 
Correspondingly, out of criticality and above $\alpha = 1$, 
we obtained no deviations from the so-called area-law ($c^{\mathrm{(VNE)}} = 0$), 
valid for short-range gapped systems, while a logarithmic deviation (with $c^{\mathrm{(VNE)}} = \frac{1}{2}$) 
is derived for $\alpha<1$. Our results are in agreement with the lattice results in 
Refs.~\cite{nostro,paperdouble,ares} and confirm the presence of new 
phases at $\alpha <1$, as inferred by the effective theory (see Section \ref{ET}).
 
\item Section \ref{crexp}: At every value of the decay exponent $\alpha$ of the pairing in (\ref{Ham}), 
the critical exponents (using the standard notation \cite{huang}) 
$\alpha$ (in classical short-range models related of the specific heat, 
not to be confused with decay exponent for the pairing of the Kitaev chain considered here) and $\beta$  (referring to the Ising order parameter) 
close to $\mu = 1$ assume the same values as for the short-range Ising (Kitaev) model. 
This fact is allowed by the simultaneous presence of and $S_{\mathrm{D}}$ and $S_{\mathrm{AN}}$  out of criticality.

\end{itemize}

In Section \ref{muminus} the critical line $\mu = -1$ and $\alpha >1$  
(for $\alpha<1$ the Hamiltonian~\eqref{Ham} acquires a mass gap) is also analyzed for the sake of comparison.
Since close to this line the minimum of the quasiparticle energy is now at $k = \pi$,  the effective action 
is composed by a single term only, unlike the situation around the line $\mu = 1$.  
The consequences of this fact on the two-points correlation functions, on the breakdown of conformal invariance, on the violation of the area-law for the von Neumann entropy and on the critical exponents for the Ising order parameter are discussed. 

In the following we will obtain and describe in detail the results mentioned above. 
Conclusions and perspectives are discussed in Section \ref{concl}, while more technical material 
is presented in the Appendices.


\section{The model}
\label{model}

We start from the Kitaev Hamiltonian with long-range pairing \cite{nostro} defined 
on a $1\mathrm{D}$ lattice:
\beq
\begin{aligned}
H_{\mathrm{lat}} = - \omega \sum_{j=1}^{L} \left(a^\dagger_j a_{j+1} + \mathrm{h.c.}\right)  - \mu \sum_{j=1}^L \left(n_j - \frac{1}{2}\right) 
+ \frac{\Delta}{2} \sum_{j=1}^L \,\sum_{\ell=1}^{L-1} d_\ell^{-\alpha} \left( a_j a_{j+\ell} + a^\dagger_{j+\ell} a^\dagger_{j}\right).
\label{Ham}
\end{aligned}
\eeq 
In Eq.~\eqref{Ham}, $a_j$ is the operator destroying a (spinless) fermion in the 
site $j=1,\cdots,L$, being $L$ the number of sites of the chain. 
For a closed chain, we define $d_\ell = \ell$ ($d_\ell = L-\ell$) if $\ell < L/2$ ($\ell > L/2$) and 
we choose antiperiodic boundary conditions \cite{nostro}. 
We measure energies in units of $2 \omega$ and lengths in units of the lattice spacing $s$. 
We also set for simplicity $\Delta=2 \omega$, since as discussed in \cite{nostro} the critical values 
of the chemical potential $\mu=\pm 1$ does not depend on $\Delta/\omega$.

The spectrum of excitations is obtained via a Bogoliubov transformation and 
it is given by
\begin{equation}
\lambda_{\alpha}(k) = \sqrt{\left(\mu - \cos{k} \right)^2 + f_{\alpha}^2(k + \pi)}.
\label{eigenv}
\end{equation}
In Eq.~\eqref{eigenv}, $k=- \pi + 2\pi \left(n +  1/2\right)/L$ 
with $0 \leq n< L$ and
$f_{\alpha} (k) \equiv \sum_{l=1}^{L-1} \sin(k l)/d_\ell^\alpha$. 
The functions $f_\alpha(k)$ can be also evaluated in the thermodynamic limit, 
where they become polylogarithmic functions \cite{grad,abr,nist}. 

The spectrum Eq.~\eqref{eigenv} displays the critical point $\mu = 1$ 
for every $\alpha$ and the critical point $\mu = -1$ for $\alpha >1$.   
The ground state of Eq.~\eqref{Ham} is 
given by $|{\mathrm{GS}} \rangle =\prod_{n=0}^{L/2-1} 
\left(\cos\theta_{k} - i \sin\theta_{k} a^\dagger_{k} 
a^\dagger_{-k} \right) |0\rangle$, 
with $\tan(2\theta_{k}) = -f_{\alpha}(k+ \pi)/(\mu -\cos{k} )$, {while 
the ground state energy density $e_0(\alpha, L)$ is given by 
$e_0(\alpha, L)=-\sum_k \lambda_{\alpha}(k)/(2L)$. We remind that no Kac rescaling 
\cite{libro} is needed for the Kitaev Hamiltonian of Eq.~\eqref{Ham}, 
since $e_0(\alpha, L)$ remains finite in the $L \to \infty $ limit (see also Appendix \ref{scaling_GS}).


\section{Critical effective theory}
\label{ET}

In this Section  we derive an effective theory for the long-range paired Kitaev chain Eq.~\eqref{Ham} 
by using a RG approach, valid  close to the critical lines $\mu = \pm 1$. 

First we focus on a region in proximity of the critical line $\mu = 1$, where  
the minimum of the spectrum in Eq.~\eqref{eigenv} is at $k = 0$ (later in the text we will discuss 
the case $\mu=-1$). For $L \to \infty$ the Hamiltonian of Eq.~\eqref{Ham} reads 
\beq
H = \frac{L}{2\pi} \,   \int_{-\pi}^{\pi} \, \m{d} k \, \, \psi^{\dagger} (k)  \, h_{\alpha} (k) \, \psi (k), 
\label{Ham2}
\eeq
where $h_{\alpha} (k) = - (\mu - \cos{k} ) \, \sigma_3 - 
\Delta \, f_{\alpha} (k + \pi) \, \sigma_2$
acts on the space of Nambu spinors $\psi(k)=(a_{k} , a^\dagger_{-k})^T$
and the $\sigma_i$ are the Pauli matrices. 

We now proceed to the RG procedure, which for the long-range Kitaev chain 
takes advantage of the quadratic nature of the model. 
We refer to Ref.~\cite{continentino} for an exact RG treatment of the short-range Kitaev chain. 
Starting from a theory having an energy cut-off $\Lambda$ 
(here equal to $\pi$), the standard strategy divides in the following three steps \cite{huang,shankar}: 
{\em a) decimation:} integrate out high-energy portions of the momentum space, that is 
in our model between $\pm \pi$ and $\pm \frac{\pi}{b}$ (with $b>1$). The resulting Hamiltonian is 
denoted by $H_{\mathrm{L}}^{(b)}$; 
{\em b) rescaling:} restore the old integration domain for the momenta, 
by redefining $k^{\prime}= b \, k$;
{\em c) renormalization:} reabsorb the effect of $b$ in the normalizations of the 
fields and in the parameters appearing in $H_{\mathrm{L}}^{(b)}$.

The standard prescription for decimation described above would lead to write:
\beq
H_{\mathrm{L}}^{(b)} = \frac{L}{2\pi} \, \int_{-\frac{\pi}{b}}^{\frac{\pi}{b}} \m{d} k \, 
\psi^{\dagger} (k) \, h_{\alpha} (k) \, \psi (k),
\label{hRG}
\eeq
after integrating out the momenta far from the minimum of the spectrum at $k = 0$. 
However decimation must be performed with particular care for Eq.~\eqref{Ham}. Indeed one should not discard
the contributions of the high-energy modes at the edges of the Brillouin zone, where singularities develop
\big(in the $[\alpha]$-th derivative of the spectrum $\lambda(k)$, $[\alpha]$ labeling the integer part of
$\alpha$\big) due to the long-range nature of the model encoded in $f_{\alpha}(k)$. 
Indeed previous works \cite{nostro,paperdouble} have shown that these
high-energy modes heavily affect various basic properties of Eq.~\eqref{Ham}. 
In particular they determine the peculiar hybrid decay of the correlation functions 
and the anomalous scaling of the ground-state energy density at criticality 
(see Section \ref{intro} and  Appendix \ref{scaling_GS}). 
Thus we proceed keeping  the contributions of the modes at the edges of the Brillouin zone during 
the decimation procedure. In Section \ref{latcorr} we verify the validity of the proposed RG procedure 
by comparing the correlation functions obtained from the effective theory with the lattice correlation functions.

This way of performing RG implies a different hierarchy for the quasiparticles weights along the RG, 
compared to the one generally assumed for short-range systems \cite{huang}. 
Alternative hierarchies have been discussed in the context of RG theory (see Ref.~\cite{janos} and 
references therein) and concerning localization in long-range systems with disorder \cite{germandis}. 
We also observe that if had one performed exactly the RG procedure, 
then the contribution to the effective theory by the modes at the edges 
of the Brillouin zone would have been automatically correctly 
reproduced: however, to perform analytically the procedure we resort to 
expansions of the energy spectrum $\lambda_\alpha(k)$ and this requires to explicitly single out the contribution 
from the edges the Brillouin zone.

Using the procedure described above, 
and absorbing the factor $\frac{L}{2 \pi}$ in the normalization of the 
fermionic field, we write instead of Eq.~\eqref{hRG}:
\begin{equation}
H_\m{L}^{(b)} = H_\m{D}^{(b)} + H_\m{AN}^{(b ; \alpha)} 
\label{H2}
\end{equation}
with
\begin{equation}
H_\m{D}^{(b)} =  \int^{\frac{\pi}{b}}_{-\frac{\pi}{b}} \, \m{d} p \, \psi^{\dagger}\left(p\right) \, h_{\alpha}(p)\, \psi\left(p\right)
\label{eqn:DiracPart}
\end{equation}
and
\begin{equation}
H_\m{AN}^{(b; \alpha)} = \int^{\frac{\pi}{b}}_{-\frac{\pi}{b}} \, \m{d} p \, \psi^{\dagger}\left(p\right) \, 
h_{\alpha} (\pi + p) \, \psi\left(p\right).
\label{eqn:AnomalousPart}
\end{equation}
Here, $h_{\alpha}(p)$ and $h_{\alpha}(p+\pi)$ have support close to the momentum $k= 0$ and $k = \pi$, respectively. 
We exploited the periodicity of the Brillouin zone and assumed in $H_\m{AN}^{(b; \alpha)}$
the cut-off for the $p$ momenta to be again $\frac{\pi}{b}$, with $b\gg1$. 

In order to perform the rescaling and renormalization steps it is useful to 
expand $f_{\alpha} (k)$ and $f_{\alpha} (\pi + k)$ in Eqs.~\eqref{eqn:DiracPart} and \eqref{eqn:AnomalousPart} 
in powers of $k$ (see Appendix \ref{expansion}). We obtain from Eq.~\eqref{eqn:DiracPart} 
\beq
H_{\mathrm{D}}^{(b)} =  \int^{\frac{\pi}{b}}_{-\frac{\pi}{b}}  \m{d} p \, 
\bar{\psi}_{\mathrm{L}} (p) \, \Big(v_F \,  \gamma_1  \, p  + m_0 \, v_F^2 \Big) \, \psi_{\mathrm{L}} (p),
\label{IRupper}
\eeq
where $\psi_{\mathrm{L}}(p)=\psi(p)$, $\bar{\psi}_{\mathrm{L}}(p) \equiv \psi^{\dagger}_{\mathrm{L}}(p) 
\, \gamma_0$, and $m_0 \, v_F^2 \propto  |\mu-1|$ vanishes at criticality (the subscript $_0$ on $m_0$ and on the constants in the following denotes bare quantities, 
and it is removed for the corresponding renormalized quantities).  
Moreover $v_F \equiv 1$ is the rescaled Fermi velocity in $k = 0$ and,  
following the specific form of the tight-binding matrix from Eq.~\eqref{Ham}, 
we conventionally choose $\gamma_0 = -\sigma_3$ and $\gamma_1 = -i \, \sigma_1$ (see Appendix \ref{soluzioni}).

Equation \eqref{IRupper} is the usual Dirac Hamiltonian, describing, as it is well known, 
the short-range Ising model at 
criticality \cite{muss}, and it provides the dynamics around the minimum at $k = 0$. 
A central point of this Section is that the contribution $H_{\mathrm{AN}}^{(b ; \alpha)}$ takes instead a different form 
for $\alpha>2$ and $\alpha<2$. For $\alpha>2$ one has
\begin{equation}
H_{\mathrm{AN}}^{(b; \alpha>2)}=  \int^{\frac{\pi}{b}}_{-\frac{\pi}{b}}  \m{d} p \, 
\bar{\psi}_{\mathrm{H}} (p) \, \Big[  \gamma_1 \, \big(c_{1,0} \, p \,+ {\, c_{3,0} \, p^3}  + \dots + a_0 \, p^{\beta} \big) \, + \, 
M_0 \Big] \, \psi_{\mathrm{H}} (p),
\label{IRlower2}
\end{equation}
where $\psi_{\mathrm{H}}(p)=\psi(\pi+p)$ and $M_0 (\mu)  \equiv | \mu + 1|$. The coefficient 
$\beta$ (not to be confused with the scaling exponent $\beta$ mentioned in the Introduction) 
is given by 
\begin{equation}
\beta \equiv \alpha -1.
\label{beta_def}
\end{equation} 
In Eq.~\eqref{IRlower2} the coefficients $a_0$ and $c_{n,0}$ ($n \ge 1$) 
have the expressions given in Eq.~\eqref{eqn:TaylorOfEff}. These expressions are $\alpha$-dependent, 
although in Eq.~\eqref{IRlower2} and in the following this dependence is dropped for the sake of brevity. 
Again in Eq.~\eqref{IRlower2} the sum over the odd $n$'s is up to the largest integer smaller than $\beta$. 
The power $p^{\beta}$ in Eqs.~\eqref{IRlower2} and \eqref{IRlower} has to 
be interpreted here and in the following as
$p^{\beta} \equiv \mathrm{sign}(p) \,  |p|^{\beta}$. 
This definition comes directly from the two different expansions in $p$-powers series of 
$f_{\alpha}(\pi + p)$ in Eq.~\eqref{H2}, depending on $\mathrm{sign}(p)$,
as discussed in Appendix \ref{expansion}.

For $\alpha < 2$ we find
\beq
H_{\mathrm{AN}}^{(b; \alpha<2)} =  \int^{\frac{\pi}{b}}_{-\frac{\pi}{b}}  \, \m{d} p \, \bar{\psi}_{\mathrm{H}}(p) \, 
\Big[a_0 \, \gamma_1 \,  p^{\beta}  + M_0 \Big] \, \psi_{\mathrm{H}} (p).
 \label{IRlower}
\eeq
The Hamiltonian~\eqref{IRupper} commutes with those in Eqs.~\eqref{IRlower2} and \eqref{IRlower}. 
This fact is at the origin of the good agreement between effective theory results and numerical 
findings not only at asymptotically large distances, but also at intermediate length scales.

Notice that while in Eq.~\eqref{IRupper} and Eq.~\eqref{IRlower} we retained 
only the leading term in the expansion of $f_{\alpha} (\pi + p)$ in $p$-powers, 
in \eqref{IRlower2} also subdominant terms are kept for future convenience. 
Since these terms are suppressed along the RG flow  they will be discarded in the following of the present Section, 
where RG is analyzed. As discussed in textbooks \cite{dif,muss} $\psi_{\mathrm{L}}(p)$ and $\psi_{\mathrm{H}}(p)$ 
are Majorana fields \cite{majorana,pal}; 
their canonical quantization is described in Appendix \ref{soluzioni}. \\

Rescaling and renormalization [points {\em b)} and {\em c)}] in Eq.~\eqref{H2} proceed now as follow.
Under rescaling $p \to \frac{p^{\prime}}{b}$, $H_{\mathrm{D}}^{(b)}$ 
transforms as (omitting the primes) 
\beq
H_{\mathrm{D}}^{(b)} =  \int^{\pi}_{-\pi} \frac{\m{d} p}{b} \, \bar{\psi}_{\mathrm{L}} \Big(\frac{p}{b}\Big) \, \Big(  \gamma_1 \, \frac{p}{b} + m_0  \Big) \, \psi_{\mathrm{L}} \Big(\frac{p}{b}\Big),
\eeq
while $H_{\mathrm{AN}}^{(b)}$ transforms as 
\beq
H_{\mathrm{AN}}^{(b; \alpha>2)} \approx  \int^{\pi}_{-\pi} \frac{\m{d} p}{b} \, 
\bar{\psi}_{\mathrm{H}} \Big(\frac{p}{b}\Big) \, \Big[c_{1,0} \, \gamma_1 \, \frac{p}{b} + M_0 \Big] \, \psi_{\mathrm{H}} \Big(\frac{p}{b}\Big),
\eeq
for $\alpha>2$ and as 
\beq
H_{\mathrm{AN}}^{(b; \alpha<2)} =  \int^{\pi}_{-\pi} \, \frac{\m{d} p}{b} \, 
\bar{\psi}_{\mathrm{H}} \Big(\frac{p}{b}\Big) \, \Big[a_0 \,  \gamma_1 \,\frac{p^{\beta}}{b^{\beta}}  + M_0 \Big]
  \, \psi_{\mathrm{H}} \Big(\frac{p}{b}\Big)
\eeq
for $\alpha<2$.
Standard power-counting arguments \cite{peskin,wei1} 
show that the field $\psi_{L}(p)$ has dimension 
$d_{\psi} = \frac{1}{2}$ in mass (with $\hbar = v_F = 1$). Therefore, 
under rescaling it transforms as 
$\psi_{L}(\frac{p}{b}) = \psi_{L}(p)\,  b^{\frac{1}{2}}$. 
The same scaling law applies to $\psi_{\mathrm{H}}(p)$, since 
the mass terms in Eqs.~\eqref{IRupper} and \eqref{IRlower} 
have the same functional form. 
Equivalently, $\psi_{\mathrm{H}}(p)$ differs from $\psi_{L}(p)$ 
only for the lattice momentum $k$ where $p$ is centered around ($k= \pm \pi$ and  $k = 0$ respectively). 

With these scaling laws we obtain:
\beq
H_{\mathrm{D}}^{(b)} =  \int^{\pi}_{-\pi} \m{d} p \, 
\bar{\psi}_{\mathrm{L}} (p) \, \Big(\gamma_1 \, \frac{p}{b} + m_0 \Big) \, 
\psi_{\mathrm{L}} (p),
\label{IRupper3}
\eeq
(similarly for $ H_{\mathrm{AN}}^{(b; \alpha>2)}$) and
\beq
H_{\mathrm{AN}}^{(b; \alpha<2)} =  \int^{\pi}_{-\pi} \, \m{d} p \, \bar{\psi}_{\mathrm{H}}(p) \,\Big[ a_0 \,  \gamma_1 \, \frac{p^{\beta}}{b^{\beta}}  + M_0 \Big] \, \psi_{\mathrm{H}} (p).
 \label{IRlower3}
\eeq
In Eq.~\eqref{IRupper3} the $b$-dependence can be reabsorbed 
by the renormalization transformation (not related with the rescaling law) 
$\psi_{\mathrm{L}} (p) \to \psi_{\mathrm{L}} (p) \, b^{\frac{1}{2}}$, 
also implying $m_0 \to b \, m_0 \equiv m$. { Similarly for $M_0$ in  $ H_{\mathrm{AN}}^{(b; \alpha>2)}$
we find: $M_0 \to b \, M_0 \equiv M$.} 
Assuming again for $\psi_{\mathrm{H}}(p)$ the same renormalization as for $\psi_{\mathrm{L}}(p)$, 
we obtain for Eq.~\eqref{IRlower3}: 
$a_0 \to \frac{a_0}{b^{(\beta-1)}} \equiv a$ and $M_0 \to b \, M_0 \equiv M$.
A different choice for the renormalization of $\psi_{\mathrm{H}}(p)$ does not change 
the evolution along RG of the ratio between the renormalized parameters $a$ and $M$.

In order to better analyze the RG flow of Eq.~\eqref{H2}, we discuss
separately the cases $\alpha >2$ and $\alpha< 2$:
\begin{itemize}

\item If $\alpha >2$, after neglecting the subdominant $p$-powers  in Eq.~\eqref{IRlower2},
both Eqs.~\eqref{IRupper} and \eqref{IRlower2} become 
Dirac Hamiltonians. If $\mu = 1$, then $m=0$ for Eq.~\eqref{IRupper}, while $M \neq 0$ for Eq.~\eqref{IRlower2}. Since 
these masses renormalize with the same factor $b$, we conclude then that $H_\mathrm{D}$ is 
dominant against $H_{\mathrm{AN}}^{(\alpha>2)}$ along the RG flow, as an effect of the mass term $\propto M \gg m $. 
The role played by the modes at $k  \approx  \pm \pi$ is suppressed for the dynamics close 
to criticality in this range of $\alpha$. One thus retrieves the Majorana field theory proper of the critical short-range Ising model.

\item If $\alpha <2$ a direct comparison between the $p$-powers of their kinetic terms 
shows that  $H_{\mathrm{AN}}^{(\alpha<2)}$ becomes dominant with respect 
to $H_\mathrm{D}$ along the RG flow. Thus, due to the contributions at $k \approx \pm \pi$, 
the RG flow leads to a critical theory different from 
the Dirac one predicted around the minimum of the lattice spectrum. 
The analysis of the effective theory shows that one may further distinguish the case 
$1<\alpha< 2$ and $\alpha<1$:
\begin{itemize} 

\item For $1<\alpha<2$ the ratio $\frac{M}{a}$ increases along the flow: 
fixing conventionally $a = a_0$, one finds in this unit that $M \to b^{\beta} M$, with $\beta >0$. 
More precisely, 
since $M_0(\mu = 1)  \neq 0$, we obtain that $M$ stays nonzero also on the critical line $\mu = 1$, 
which is massless for the lattice spectrum. The term $\propto M$ produces important effects 
on the dynamics close to criticality: as discussed in Section \ref{corrET}, 
it allows for an asymptotic decay of correlations functions consistent with lattice calculations (Section \ref{latcorrsub}). 
In particular, although $M \neq 0$, the decay of static correlation functions is still algebraic and with exponents in agreement with the ones derived in Section \ref{latcorr}. We observe that if one 
sets $M \to 0$ a wrong decay would rather be obtained, providing a check for the 
soundness of our RG scheme.

\item If $\alpha <1$, then $\beta < 0$ and $M$ tends to vanish along the RG. 
This behavior for $M$, different from the one at $\alpha >1$, is a signature of a  
quantum phase transition arising along {a line passing through $\alpha  = 1$ at $\mu = 1$}, as 
argued in \cite{nostro}. 

\end{itemize}

\end{itemize}

The RG flow of $M$ in the three different ranges for $\alpha$ is plotted in Fig.~\ref{masse}.   
\begin{figure}[h!]
\includegraphics[scale=0.5]{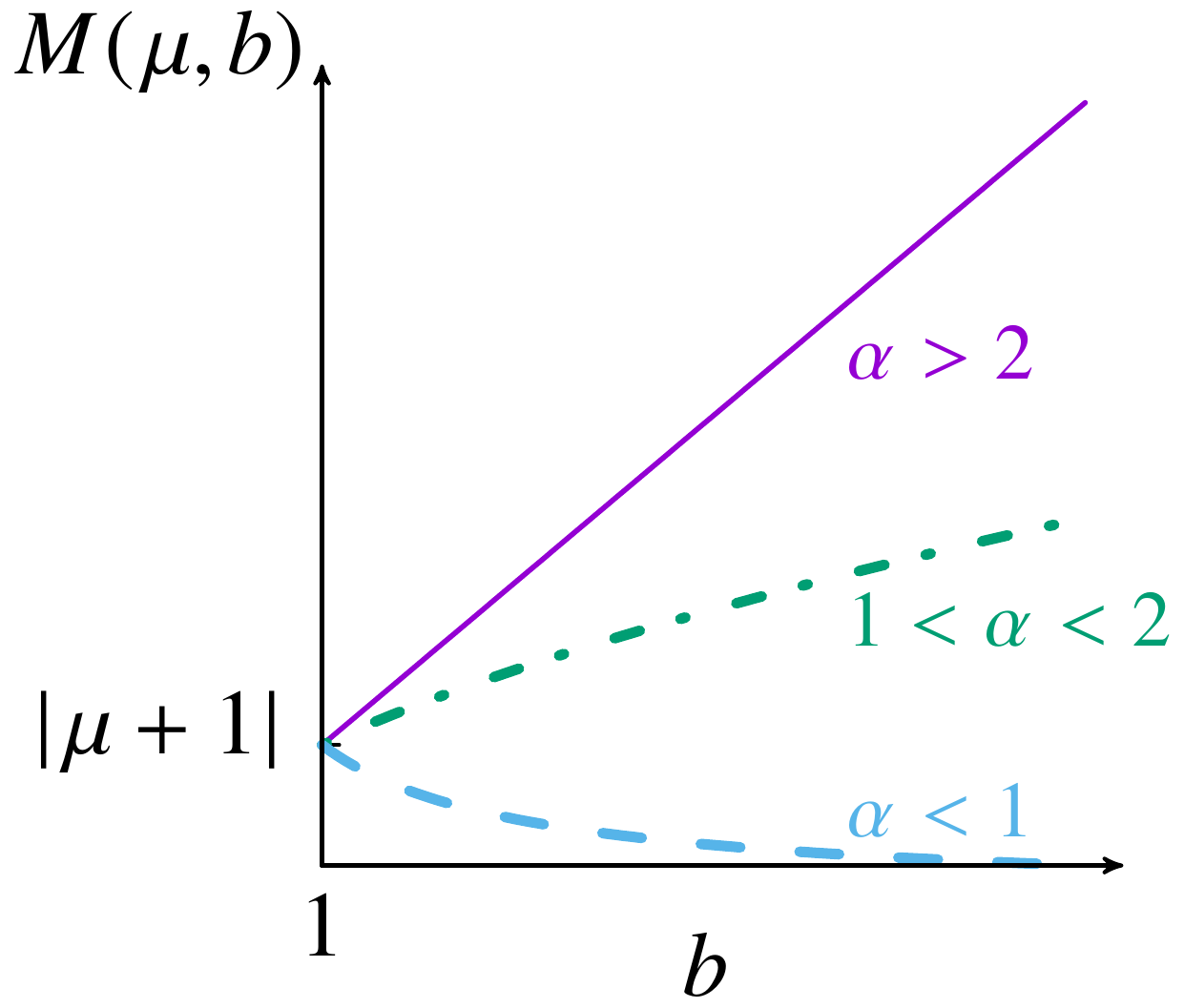}
\caption{Typical evolution for $M$ under RG flow 
(encoded in the dependence on the decimation scale $b$) for $\alpha >2$, $1<\alpha<2$, and $\alpha <1$. 
The origin of the $\hat{x}$ axes is in $b = 1$, where $M(\mu, 1) = M_0 = |\mu+1|.$}
\label{masse}
\end{figure}

We notice that Hamiltonians (or corresponding actions) with non integer $p$-exponents, 
similar to Eqs.~\eqref{IRlower2}-\eqref{IRlower}, 
have been presented in literature as effective theories of long-range models (see e.g. 
Refs.~\cite{defenu,Maghrebi2015} and references therein). However to our knowledge 
the example in this paper is the first one 
in which is obtained by RG from microscopics for a quantum long-range model.

Finally, in one dimensional quantum systems with continuous symmetries a RG 
behavior as the one described in the present Section is expected related with 
the breakdown of the Mermin-Wagner theorem \cite{MW,lebellac}, generally occurring 
at small enough values of $\alpha$ \cite{dyson1969,thouless1969,anderson1970,cardy1981,frolich1982,lui2001,lui1997,Magh_bis}.


\section{Two-points correlation functions} 
\label{latcorr}

In this Section we compare the two-fermions correlation functions computed from the effective theory close to the 
critical line $\mu = 1$ with the ones directly computed on the lattice model.

\subsection{Lattice results}
\label{latcorrsub}

The lattice Hamiltonian in Eq. (\ref{Ham}), being quadratic, allows for 
a straightforward computation of the correlation functions. We discuss 
in the following the behavior of the lattice correlation 
$g_1^{\text{(lat)}}(R) \equiv \langle a^\dag_R a_0 \rangle$, referring to~\cite{nostro,paperdouble} and 
Appendix \ref{appcorr} for more details. Other correlation functions, as density-density ones, 
can be built from $g_1^{\text{(lat)}}(R)$ and 
$g_1^{\text{(lat; anom)}}(R) \equiv \langle a^\dag_R a^\dag_0\rangle$ (also calculated in Appendix \ref{appcorr})
by Wick's theorem.

In the limit $L\to\infty$, $g_1^{\text{(lat)}}(R)$ reads
\begin{equation}
g_1^{\text{(lat)}}(R) = - \frac{1}{2\pi} \int_{-\pi}^{\pi} \m{d} k \,
e^{i k R} \, \mathcal{G}_\alpha(k),
\label{formcorr} 
\end{equation}
with
$\mathcal{G}_\alpha(k)=\frac{\mu - \cos{k}}{2\lambda_\alpha(k  + \pi)}$. 
We focus on the range $\mu >0$ and at the beginning outside of the critical line $\mu = 1$. There the explicit calculation of the integral in Eq.~\eqref{formcorr} gives
\begin{equation}
g_1^{\text{(lat)}}(R) = A(R, \xi_1,\xi_2) +  B(R)
\label{corr}
\end{equation}
with $A(R, \xi_1,\xi_2) = (-1)^R  \, \mathcal{D}_\alpha(\xi_1,\xi_2) \,  e^{- \xi_1 R}$ and
\begin{itemize}
\item[*)] $B(R) =  - \frac{2\alpha}{(\mu+1)^2} \frac{\zeta(\alpha-1)}{R^{\alpha+1}}$  if $\alpha > 2$; 
\item[**)] $B(R) =  \frac{\cos^2(\pi\alpha/2) \sin (\pi\alpha)}{\pi(\mu+1)^2}\frac{\Gamma(2\alpha-1)}{R^{2\alpha-1}} $ if $ 1<\alpha<2$; 
\item[***)] $B(R) = \frac{\mu+1}{4\pi} \frac{1-\alpha}{R^{2-\alpha}} $ if $ 0 < \alpha < 1$. 
\end{itemize}
The parameters $\xi_1$ and $\xi_2$ are zeros of 
$\lambda_{\alpha}(z)$ and depend on $\mu$ and $\alpha$ 
in an implicit way.  $\Gamma(x)$ and 
$\zeta (x)$ denote the Gamma and the Riemann zeta functions, respectively. 
The expression $\mathcal{D}_\alpha(\xi_1,\xi_2)$ 
is just a multiplicative constant whose expression can be extracted from the results 
of Appendix \ref{appcorr}.

The first contribution $A(R, \xi_1,\xi_2)$ in Eq.~\eqref{corr} decays purely exponentially 
outside of the massless line and 
it is due to momenta close to the minimum of the energy, at $k \approx 0$. 
The second contribution $B(R)$ decays instead algebraically  
and it originates from the higher energy momenta around $k = \pm \pi$. 
When the two contributions become of the same order, the change of decay 
from exponential to algebraic takes place \cite{cirac05,paperdouble}. 
We notice that the algebraic tail occurs in the presence of a nonzero lattice mass 
gap, as found also in  \cite{cirac05,tagliais,nostro}. 
This hybrid behavior has also been observed in the Ising model with long-range interactions 
\cite{cirac05, paperdouble} and seems to be 
general for long-range systems.

Since it will be useful in the following, we write down 
the explicit expression for $B(R)$: 
\begin{equation}
B(R) = \frac{1}{\pi}  \int_{0}^\infty  \m{d} p \, e^{- p R} \, \mathrm{Im}  \big[\mathcal{G}_{\alpha}(i  p) \big].
\label{integrpi}
\end{equation}
The asymptotical behavior of Eq.~\eqref{integrpi} can be calculated by integrating the 
main contribution of $\mathcal{G}_{\alpha}(i  p)$ in the limit $p\to 0$ 
\cite{Ablowitz2003}.
This term can be evaluated exploiting a series expansion 
of $f_{\alpha}(k)$ entering in $\lambda_{\alpha}(k)$, as in Section \ref{ET}. The result is 
\beq
\mathcal{G}_\alpha(p) \sim \frac{M_0}{\sqrt{(\mu+1)^2 +r_2 \,  
p^2+ r_{2\alpha-2} \, p^{2\alpha-2} + r_{\alpha} \, p^{\alpha} + \dots}}, 
\label{gfun}
\eeq
where $r_{2\alpha-2}$ and $r_{\alpha}$ are complex factors, while $r_2$ is real: their expression can be deduced 
from the expansion of $f_\alpha^2(p+\pi)$ given in Appendix \ref{expansion}. 
In this way the term $\propto p^2$ does not contribute to the imaginary part 
of $\mathcal{G}_\alpha(i p)$ in Eq.~\eqref{integrpi}. In Eq.~\eqref{gfun} 
the symbol $\dots$ labels 
terms with integer $p$-powers larger than $2$, 
not contributing to  $\mathrm{Im} \, \mathcal{G}_\alpha(i p)$, as well as the higher non integer powers, 
whose contribution to $\mathrm{Im} \, \mathcal{G}_\alpha(i p)$ is suppressed in  the limit $p \to 0$. 
Notice also that if $\alpha < 2$ the leading term for $p \to 0$ in 
the denominator of Eq.~\eqref{gfun} is the one $\propto p^{2 \beta}$ 
(with again $\beta = \alpha -1$).

Exactly at criticality a similar calculation starting from Eq.~\eqref{formcorr} yields:
\begin{equation}
g_{1,\mathrm{cr}}^{\text{(lat)}}(R) = A_{\mathrm{cr}}(R) +  B_{\mathrm{cr}}(R)
\label{corrcrit}
\end{equation}
with $A_{\mathrm{cr}}(R) = \frac{A(\alpha)}{R}$ (this decay, due to the modes close to the zero point of the energy spectrum, is expected, being the same as for the short-range Ising model at criticality) and $B_{\mathrm{cr}}(R) $ trivially the same as in Eqs.~\eqref{corr}, points $*), **), ***)$, with $\mu = 1$.
It is important to notice that for $\alpha >1$
the dominating term for $g_{1,\mathrm{cr}}^{\text{(lat)}}(R)$ in the limit $R \to \infty$ is $A_{\mathrm{cr}}(R)$, while for $\alpha <1$ both $A_{\mathrm{cr}}(R)$ and $B_{\mathrm{cr}}(R)$ decay as $\frac{1}{R}$ in the same limit.


\subsection{Effective theory results}
\label{corrET}

The analysis performed in Section \ref{ET} can be summarized as follows. Close to the critical 
point $\mu = 1$, the effective theory can be described as the sum of two actions:   
\beq
S= S_{\mathrm{D}}+  S_{\mathrm{AN}},
\label{aztot}
\eeq
with $S_{\mathrm{D}}$ the Euclidean Dirac action corresponding to Eq.~\eqref{IRupper} and $S_{\mathrm{AN}}$ 
the actions corresponding  to
Eqs.~\eqref{IRlower2} and \eqref{IRlower} for $\alpha>2$ and $\alpha<2$, respectively. 
For $\alpha<2$ the Euclidean action corresponding to Eq.~\eqref{IRlower} reads 
(we set $a \equiv 1$ after global redefinition for the fermionic fields and without any loss of generality):
\beq
S_{\mathrm{AN}}^{(\alpha<2)} = 
\int  \m{d} x  \, \m{d}\tau \, \bar{\psi}_{\mathrm{H}} (\tau,x) \,  \Big[\gamma_0 \, \partial_\tau \,  + \,  \gamma_1 \,  \partial_x^{\beta}  \, + \,M  \Big] \,  \psi_{\mathrm{H}} (\tau,x).
\label{lren3}
\eeq
In order to extend the interval of integration in Eq.~\eqref{IRlower3} to the whole real line, as in 
Eq.~\eqref{lren3}, we reintroduced
the lattice spacing $s$ by replacing $\pi$ with $\frac{\pi}{s}$, and 
then took the limit $s \to 0$. The notation using the fractional derivative means 
that the inverse propagator of the effective action in Fourier space depends on $p^\beta$, 
as customarily done in the treatment of long-range systems \cite{defenu}.

In this Subsection we calculate the correlation $g_{1}^{(\mathrm{ET})} (R) \equiv \langle a^\dag_R a_0 \rangle$ 
around $\mu = 1$ using the effective theory in Eq.~\eqref{aztot}, and compare it to the lattice results 
previously given.
From the effective theory we expect 
to reproduce the large separation behavior of the corresponding lattice correlations.

Depending on $\alpha$, $S_{\mathrm{D}}$ and $S_{\mathrm{AN}}$ become dominant along the RG flow 
respectively for $\alpha>2$ and $\alpha<2$. 
However the two terms in the action do not decouple completely, at least out of criticality, 
and they co-operate to determine all the dynamical quantities 
(as correlations or entanglement properties). In other words, the correlations 
have to be computed using the effective theory renormalized at the momentum scale 
$\frac{\pi}{b}$. Only afterwards the quantity $b$ can be sent to infinite.

Since the actions in Eq.~\eqref{aztot} are commuting, $g_{1}^{(\mathrm{ET})} (R)$ is composed again by the sum of 
two contributions. Let us start from the non critical correlation functions. 
The first contribution from $S_{\mathrm{D}}$ is well known \cite{muss} 
to have in the massive regime the same decay (exponential) of the 
lattice term $A(R, \xi_1,\xi_2)$ (see Subsection \ref{latcorr}). The second one from $S_{\mathrm{AN}}$ 
can be evaluated by the propagator in the Minkowski space-time corresponding to the action (\ref{lren3}):
\beq
\tilde{D}(p_0, p) = \frac{1}{p_0 \,  \gamma_0  -  \,  p^{\beta} \, \gamma_1 - M}
\label{corr_D}
\eeq 
(as in \eqref{lren3}, we set $p^{\beta} \equiv \mathrm{sign}(p)  |p|^{\beta} $). 
Multiplying by $\Big(p_0 \,  \gamma_0  - \,  p^{\beta} \, \gamma_1 + M \Big) $ 
both the numerator and the denominator of Eq.~\eqref{corr_D}, 
exploiting the standard residue technique to perform the integration on $p_0 = \sqrt{(p^{\beta})^2 + M^2}$ 
and following 
the usual Feynman $\epsilon$-prescription \cite{peskin}, we obtain the time-ordered correlation 
\begin{equation}
\langle 0| T \, \psi_{\mathrm{H}}(x^{\mu}) \bar{\psi}_{\mathrm{H}}(y^{\mu}) |0\rangle=  
\int  \frac{d p}{4 \pi} \, e^{i   p R} \,  \Bigg[ 
 \frac{M}{\sqrt{(p^{\beta})^2 + M^2 }} 
- \gamma_1   \, \frac{p^{\beta}}{\sqrt{(p^{\beta})^2 + M^2 }}  + \, \gamma_0   \Bigg] \big[f_+ (p, t) + f_- (p,t)\big]
\label{propnto}
\end{equation}
with $x^{\mu} - y^{\mu} \equiv r^{\mu} = (t, R)$ 
and $f_{\pm} (p,t) = \theta (\pm t) \, e^{i \, \sqrt{(p^{\beta})^2 + M^2} \, t}$. 
The symbol $| 0\rangle$ denotes the ground state of the Hamiltonian from $S_{\mathrm{AN}}$.

We first focus on the two point static correlation function obtained setting $t = 0$ in Eq.~\eqref{propnto}. 
In this case, as discussed in Appendix \ref{soluzioni}, the matrix propagator Eq.~\eqref{propnto} is 
the continuum equivalent of the lattice correlation matrix:
$\begin{pmatrix} -\langle a_i a^{\dagger}_j \rangle \, \, \, \langle a^{\dagger}_i a^{\dagger}_j \rangle  \\ 
- \langle a_i a_j \rangle \, \, \,  \langle a^{\dagger}_i \, a_j \rangle 
\end{pmatrix}$.

Our goal is to evaluate the limit $R \to \infty$ of Eq.~\eqref{propnto} and compare it with the 
large distance behavior of the lattice correlations $g_1^{\text{(lat)}}(R)$ and $g_1^{\text{(lat; anom)}}(R)$ 
found in the previous Subsection and in Appendix \ref{appcorr}. 
In the limit $R \to \infty$, if $1 < \alpha < 2$ ($\beta > 0$),
the dominant part of Eq.~\eqref{propnto} is
\beq
g_{1}^{(\mathrm{ET}; \mathrm{AN})} (R) = - \frac{1}{2 \pi}  \int  
\m{d} p  \, \frac{M}{2 \sqrt{(p^{\beta})^2 + M^2}} \, e^{i p R},
\label{DOMINANT}
\eeq
since the term proportional to $p^\beta$, involving $R$-derivatives of Eq.~\eqref{DOMINANT}, gives a 
next-to-leading-order contribution. This latter term corresponds to 
the anomalous correlation function $g_{1}^{\text{(lat; anom)}} (R)$, while Eq.~\eqref{DOMINANT} corresponds 
to $g_{1}^{\text{(lat)}} (R)$.
The term in Eq.~\eqref{DOMINANT} is the effective equivalent of Eq.~\eqref{formcorr} in the continuous space limit 
$s \to 0$ and $L \to \infty$, as clear from the matrix structure of the propagator in Eq.~\eqref{corr_D}, 
and with $M$ from the renormalization of $M_0$. By the substitution $p \to i p$ one gets
\beq
g_{1}^{(\mathrm{ET};\mathrm{AN})} (R) = \frac{1}{\pi} \,  \int_{0}^\infty  \m{d} p  \,  
e^{- p  R} \, \mathrm{Im} \Bigg(\frac{M}{2\sqrt{\big((i \, p)^{\beta} \big)^2 + M^2}} \Bigg),
\label{propnto1}
\eeq
no additional term occurring due to the analytical continuation.
The imaginary factor in Eq.~\eqref{propnto1} is matching the asymptotical limit 
for $p \to 0$ of $\mathrm{Im} \big(\mathcal{G}_{\alpha}(i  p) \big)$ in Eq.~\eqref{integrpi}. 
Repeating the same calculations as for the lattice contribution 
$B(R)$ in Eq.~\eqref{corr} (see Appendix \ref{appcorr}), it is easy to show that Eq.~\eqref{propnto1} 
reproduces the power-law behavior for $B(R)$ 
denoted above by $**)$. This result confirms the correctness of Eq.~\eqref{lren3} and the role 
of the high-energy modes at the edges of Brillouin zone for the critical dynamics of the Hamiltonian (\ref{Ham}).

If $\alpha < 1 $ the first two terms in Eq.~\eqref{propnto} map 
into each others after the 
transformation $M \to \frac{1}{M}$.
Repeating the 
same calculation done for $\alpha > 1$, we have that  also in this case
$g_{1}^{(\mathrm{ET};\mathrm{AN})} (R)$ reproduces  the lattice decay behavior $***)$. 
Notice that, since in this range the energy spectrum diverges for $p \to 0$, the theory (\ref{lren3}) is well defined 
only since this divergence is integrable, 
a fact closely related to the lack of necessity for the Kac rescaling on the lattice (see Section \ref{model}). 

The same asymptotical decays $**)$ and $***)$ can be obtained as well directly 
from the first term in Eq.~\eqref{propnto}, taking respectively the limits 
$M \to \infty$ and $M \to 0$ 
and integrating the leading terms of the resulting expansions (see Appendix \ref{altra_der}). Summing up, the obtained 
agreement between the lattice correlations in Eq.~\eqref{corr} and the
ones from the effective theory in Eqs.~\eqref{aztot} and \eqref{lren3}
is a check of the correctness of Eqs.~\eqref{aztot} and \eqref{lren3} themselves.

For $ \alpha > 2$, from Eq.~\eqref{IRlower2} we find that the dominant part of $g_{1}^{(\mathrm{ET};\mathrm{AN})} (R)$ is 
\beq
g_{1}^{(\mathrm{ET};\mathrm{AN})} (R) = - \frac{1}{2 \pi} \int  \m{d} p  \, \frac{M}{ 2 
\sqrt{M^2 + 
c_2 \, p^2 + \dots + c_{\alpha} \, p^{\alpha}}} \, e^{i p R}  \, 
\equiv - \frac{1}{2 \pi} \int  \m{d} p  \, G_{\alpha}(p)  \, e^{i p R},
\label{corr+2}
\eeq 
$c_2$ and $c_{\alpha}$ being real factors and where 
the symbol $\dots$ indicates terms with even integer $p$-power exponents between  2 and $\alpha$ 
(see Formula \ref{expans2}).
The same calculation as before leads to  
an expression in agreement 
with the power-law decay $*)$ found on the lattice.

We observe that the terms with integer $p$-powers in the denominator of Eq.~\eqref{corr+2} do not 
contribute to $\mathrm{Im} \, G_{\alpha}(i p)$, 
similarly to Eqs.~\eqref{integrpi}, \eqref{gfun} and \eqref{propnto1}. Therefore 
not including in Eq.~\eqref{corr+2} the term $\propto p^{\alpha}$ 
would give rise to a second exponential tail, 
in disagreement with the lattice result $*)$. 
Thus we arrive to the remarkable conclusion that, although the terms with $p$-power exponents larger than $2$ 
in Eq.~\eqref{corr+2} are 
strongly suppressed along the RG flow approaching the critical line $\mu = 1$, their 
effects are appreciable 
in nonlocal quantities, such as correlation functions at very large separations. 
Notably to discard these terms amounts to retain only the leading order for the $p$-expansion in the Hamiltonian 
(\ref{IRlower2}).

From the analysis in the present Section, we conclude that keeping RG-subdominant terms is necessary in general to 
correctly compute nonlocal quantities of long-range systems as the two point static correlation functions. 
However a general strategy to exactly take into account  
their weights in the effective theory is still missing. For instance, keeping in the Hamiltonian (\ref{IRlower2}) 
only the terms $\propto p^2$ and $\propto p^{\alpha}$, or even only the one $\propto p^{\alpha}$, 
would have led to the same asymptotical decay for the correlation 
$g_{1}^{(\mathrm{ET};\mathrm{AN})} (R)$ above $\alpha = 2$. 
A complete solution of the problem requires more constraints on the effective theory, 
beyond the mere reproduction of asymptotical static correlation functions, or  
an exact RG treatment, as done in \cite{continentino} for the short-range limit. 
Notice that RG subleading terms do not affect the phase diagram of the model \eqref{Ham}, 
in particular the appearance of new phase(s) at $\alpha <1$. 

Agreement for the exponents of the power-law decay of the correlation functions 
between lattice results and effective theory is also found exactly at criticality ($\mu = 1$), 
where the contribution from $S_{\mathrm{D}}$  
is well known \cite{muss} to decay $\propto \frac{1}{R}$, as $A_{\mathrm{cr}}(R)$ in Subsection \ref{latcorr}. 
The contribution coming from $S_{\mathrm{AN}}$ to the correlation function $g_1^{(\mathrm{ET})}(R)$ at criticality can be 
seen to have the same exponent of $B_{\mathrm{cr}}$.

Following the same procedures described above, it is straightforward to show that 
for every $\alpha$ the decay exponents of the non-diagonal parts of the 
propagator $\langle 0| T \, \psi_{\mathrm{H}}(x^{\mu}) \bar{\psi}_{\mathrm{H}}(y^{\mu}) |0\rangle$ are in agreement with 
the ones of the anomalous correlations $g_1^{\text{(lat; anom)}}(R) \equiv \langle a^\dag_R a^\dag_0\rangle$ 
(calculated in Appendix \ref{appcorr}), both out and exactly at criticality. This confirms 
as well the correctness of our approach.


\subsection{Discussion}
\label{discussion}

In the previous Subsection we examined the large distance 
behavior of the two fermions static correlation functions computed from the effective theory \eqref{aztot}. 
The action $S_{\mathrm{AN}}$ gives both at and near criticality a power-law contribution to them, while 
Dirac action $S_{\mathrm{D}}$ gives an exponential contribution near criticality and a power-law one at criticality. 
At criticality the power-law contributions from $S_{\mathrm{AN}}$ are negligible at large distances 
for every $\alpha$, in the sense that its decay exponents are larger than the exponents coming from the Dirac action. \\
Using the total action $S=S_{\mathrm{D}}+S_{\mathrm{AN}}$ we were able to reproduce:
\begin{itemize}

\item both the exponential and power-law contributions to the correlation functions out of criticality;

\item both the  the exponential and power-law contributions to the correlations exactly at criticality.

\end{itemize}
 
A possible objection to our approach is that one could instead use only $S_{\mathrm{AN}}$ to reproduce the 
correlation functions. However, using only $S_{\mathrm{AN}}$ would have reproduced the leading decay behaviors 
at large separations only outside of the critical point, while one would have failed to reproduce the dominant 
terms at criticality, which are rather coming from $S_{\mathrm{D}}$. This implies that if one wanted to use 
only one of the two terms in $S$, he would need to choose $S_{\mathrm{AN}}$ out of criticality and $S_{\mathrm{D}}$ exactly at criticality, 
which is a rather {\em ad hoc} approach. The advantage of using both the terms in the action $S$
is then that both near and at 
criticality all the leading contributions of the correlations are reproduced in a natural way.

Another related important remark is that a careful inspection 
shows that in the range $\bar{C}= \{\alpha > \frac{3}{2} \, , \, \alpha < \frac{1}{2} \}$ 
other exponentially decaying terms develop in general from 
Eq.~\eqref{DOMINANT}, that means using $S_{\mathrm{AN}}$. An explicit 
example is given in the Appendix \ref{app:exp}. 
To ask whether the action $S_{\mathrm{AN}}$  
alone can reproduce the hybrid decay for the static correlation functions, 
at least in the range $\bar{C}$, amounts at the first level to ask 
whether such exponential terms may become algebraic at the critical point. 
One can see that that this not the case. Indeed for exponential terms from 
the Dirac action $S_{\mathrm{D}}$, the correlation length depends
on the mass $m \propto |\mu-1|$ in the same action and it diverges at the critical point. 
Conversely, the correlation length in the exponentially 
decaying terms from  Eq.~\eqref{DOMINANT} depend on $M 
\propto \mu +1$ (see Appendix \ref{app:exp}), and 
then they do not become algebraically decaying   
at criticality. 

The latter observation enforces the picture drawn in the previous Sections, involving two 
commuting actions $S_{\mathrm{D}}$ and $S_{\mathrm{AN}}$, jointly with the fact that 
out of $\bar{C}$ no exponential decaying term is found from $S_{\mathrm{AN}}$ 
(see Appendix \ref{app:exp}). Nevertheless in principle a residual possibility to 
overcome the problems described above and obtain hybrid static correlations from  $S_{\mathrm{AN}}$ alone would 
rely on a possible dependence of $S_{\mathrm{AN}}$ 
on the Dirac mass $m$, instead of on $M$. 
A similar possibility has been considered in a 
very recent work \cite{Maghrebi2015}, 
dealing with causality in long-range critical systems. 
There two dispersion contributions 
$\propto p^2$ and $\propto p^{\beta}$ 
(typically used in RG treatments of long-range systems 
\cite{cardybook,sak_1973,dutta2001, defenu}) 
are present for every $\alpha$ in an effective action with a 
single mass term, allowing for a discussion of the correlation 
functions \cite{Maghrebi2015}. 
To make a comparison with the present paper, 
we notice that in our approach, at variance, the two contributions 
$S_{\mathrm{D}}$ and $S_{\mathrm{AN}}$ (the latter one containing the term 
$\propto p^{\beta}$) derive respectively from the modes 
at the minimum of the energy spectrum and at the edges of the Brillouin zone, 
and that both the action terms have their own mass, scaling differently if $\alpha <2$.  
Assuming 
our point of view for the implementation of the RG implies that the 
choice in \cite{Maghrebi2015} 
leads to mix the dispersions of the two sets of quasiparticles. 
This may be significant for $\alpha <1$ 
(a case not treated by the authors in \cite{Maghrebi2015}), 
where the energy of the ground state is still extensive in the thermodynamic 
limit. Indeed a preliminary computation seems to indicate that extending 
the effective action introduced in 
\cite{Maghrebi2015} to $\alpha <1$ does not reproduce 
the correct area-law violation for the von Neumann entropy 
at and out of criticality (see Section \ref{VNE}). This fact
points out to the need of a systematic comparison of the two effective actions and their predictions for the 
quantities of physical interest.


\section{Breakdown of the conformal symmetry}

\label{csbreak}

The action in Eq.~\eqref{lren3}, dominant term in Eq.~\eqref{aztot} for $\alpha <2$, 
breaks explicitly the conformal group. This can be shown analyzing the behavior of Eq.~\eqref{lren3}
under the global part of the conformal group \cite{dif,muss}. Summing up the results of this analysis, 
when $1< \alpha< 2$, $\frac{M}{a}$ is increasing along the RG flow, thus
one concludes that CS is broken 
by the mass term $M$. 
Another source for the CS breakdown is the anomalous exponent for the spatial derivative in Eq.~\eqref{lren3}, 
$\beta \neq 1$, as detailed in Appendix \ref{breaking_CS}. For the same reason the CS breakdown arises also
at $\alpha < 1$, even if $\frac{M}{a} \to 0$. 

We observe that these results are consistent with our 
findings for the scaling of the ground state energy density, as discussed 
in detail in Appendix \ref{scaling_GS}: below $\alpha = 2$ the scaling law predicted by CS 
\cite{dif,muss} starts to fail \cite{nostro,paperdouble}. 

Although the action $S_{\mathrm{AN}}$ is dominant with respect to $S_{\mathrm{D}}$ for $\alpha <2$, 
a complete decoupling between the two actions cannot occur in this range, even exactly at criticality. 
This non complete decoupling is suggested by the fact that at criticality and for every $\alpha$ the leading part 
of the lattice correlation functions $g_1^{\text{(lat)}}(R)$ in the limit $R \to \infty$ is reproduced 
by the contribution from the Dirac action $S_{\mathrm{D}}$. 
Again we find that RG subleading contributions to the total effective action $S$ for the critical model affect its physical observable not only quantitatively. Conversely, for $\alpha>2$ at criticality 
$S_{\mathrm{AN}}$ flows to zero along the RG and the leading contributions to correlation functions for $R \to \infty$ 
come from $S_{\mathrm{D}}$. Nevertheless, although near criticality $S_{\mathrm{AN}}$ is dominated 
by $S_{\mathrm{D}}$, the leading contributions to the correlations come from $S_{\mathrm{AN}}$.

We also notice that below $\alpha =1$,
a new symmetry emerges for $S_{\mathrm{AN}}^{(\alpha<2)}$ as 
$x \to \lambda^{\frac{1}{\beta}} \, x$, $\tau \to \lambda \, \tau$, 
provided the corresponding transformation $\psi_{\mathrm{H}} (x^{\mu}) 
\to \psi_{\mathrm{H}} (x^{\prime \,\mu}) \, \lambda^{\frac{1}{2 \beta}}$ holds. 
Notably this scaling law for $\psi_{\mathrm{H}} (x^{\mu})$ 
still matches the one in the presence of CS 
(see Appendix \ref{breaking_CS}) 
and with the scaling analysis in Section \ref{ET}. 
The appearance of this symmetry, approximate because of the subleading presence of $S_{\mathrm{D}}$, 
is a further indication of a new phase (or two new phases on the two sides of the line $\mu = 1$) below $\alpha = 1$, 
as discussed in Section \ref{ET}.

\subsection{Breakdown of the effective Lorentz invariance}
\label{lor_br}

It is known that for a Lorentz invariant theory a two point static correlation function 
must decay exponentially in the presence of an energy gap \cite{peskin,fred,nota2}, since in this condition 
two space-like events can be correlated only exponentially \cite{peskin}. 
It is then clear that the hybrid exponential and algebraic decay of the static two points correlations 
in gapped regime and for every finite $\alpha$ described in Section \ref{latcorr}
has to be related with the breakdown  of the ELI, 
by Eq.~\eqref{lren3}, as well as by the action related to Eq.~\eqref{IRlower2}. 
This fact, likely general for long-range systems, is non-trivial, since, at variance, 
for short-range models ELI emerges for the effective theories around their critical points 
\cite{muss}. For $\alpha \to \infty$ one has that $S_{\mathrm{AN}} \to 0$ 
and the ELI is then restored in the short-range limit.

Intuitively, for our long-range model close to criticality Lorentz invariance breaks as follows. 
The Hamiltonian of the $2\mathrm{d}$ classical lattice model corresponding to Eq.~\eqref{Ham}, 
derived by means of an (inverted) transfer matrix approach \cite{muss}, 
has long-range terms in the $\hat{x}$ direction and short-range terms 
in the $\hat{\tau}$ direction. While the exact calculation to prove this fact appears 
to be difficult starting directly 
from Eq.~\eqref{Ham}, a similar conclusion can be inferred more easily analyzing the long-range Ising model
$$H = - J \, \sum_{i, l} \, \frac{\sigma_i^{(z)} \, \sigma_{i+l}^{(z)}}{l^{\alpha}} - h  \sum_{i} \, \sigma_i^{(x)}.$$ 
By a straightforward calculation, 
it is easy to check that this model translates in the classical 
$2\mathrm{d}$ system:
\beq
H = -   \sum_{i,j} \Bigg( J_{||} \, \sum_l \, \frac{S_{i,j}^{(z)} \, S_{i+l,j}^{(z)}}{l^{\alpha}} + 
J_{\perp} \, S_{i, j}^{(z)} \, S_{i, j+1}^{(z)}  \Bigg),
\label{LRIcl}
\eeq
where $S^{(z)}_{i,j}=\pm 1$ are classical variables and 
$j$ labels the $\hat{\tau}$ direction. Independently from the exact
values of $J_{||}$ and $J_{\perp}$, Eq.~\eqref{LRIcl} displays the strong anisotropy mentioned above: 
this anisotropy cannot be reabsorbed entirely along the RG flux, at least for $\alpha$ small enough, 
and thus the breakdown of rotational symmetry close to criticality may emerge. 
This fact amounts to have a mechanism to break Lorentz invariance, since 
rotations in a $2$D Euclidean space correspond to
Lorentz transformations in a  $(1+1)$ Minkowski space-time obtained from it by the Wick rotation.
It is clear that this argument is qualitative, 
and it does not fix the precise value of $\alpha$ at which Lorentz invariance breakdown 
happens in the long-range Ising models around criticality. 

Exactly at criticality CS and ELI are closely related, since
the conformal group contains, as a global subset, the Euclidean rotations \cite{dif,muss} 
(see also Appendix \ref{breaking_CS}). In spite of this fact, CS and (near criticality) 
ELI are not directly connected in general, since Lorentz invariance breaking terms 
possibly present in the non critical effective theory can vanish at criticality, due to the RG flow. 
Our model is an instance of this latter situation, since for $\alpha>2$ out of criticality the system does not have 
ELI but it does (and it has as well CS) at the critical point. Notice however that the ELI for $\alpha >2$ is broken in a "soft" way, since $S_{\mathrm{D}} \gg S_{\mathrm{AN}}$ in the RG sense. Our findings are summarized 
in Table~\ref{table1}.

\begin{table}
\begin{center}
\begin{tabular}{ |l | c |  c| }
  \hline
  $\mu=1$           & Critical point     & Near criticality \\ \hline
  $\alpha>2$ & CI and ELI         & Non ELI          \\ \hline
  $\alpha<2$ & Non CI and Non ELI & Non ELI          \\ \hline
\end{tabular}
\caption{Summary of obtained results 
as a function of $\alpha$ at and near criticality for the critical line $\mu=1$: CI stands 
for Conformal Invariance, and ELI for Effective Lorentz Invariance.}\label{table1}
\end{center}
\label{tabella}
\end{table}


\section{von Neumann entropy and area-law violation}
\label{VNE}

In this Section, we focus on the behavior close to the critical line $\mu = 1$ of the von Neumann entropy $V(\ell)$ after 
a partition of the system into two subsystems containing $\ell$ and $L-\ell$ sites respectively. 
The von Neumann entropy is defined for a $1\mathrm{D}$ chain as 
$V(\ell) = -\mathrm{Tr} \, \rho_{\ell} \log_2 \rho_{\ell}$, $\rho_{\ell}$ 
being the reduced density matrix of the subsystem with  $\ell$ sites. A study of the von Neumann entropy 
in $1\mathrm{D}$ fermionic chains with long-range couplings is given in \cite{gori_15}.

In particular, we analyze the deviations from the so-called area-law, 
that is generally valid for short-range gapped systems \cite{eisert} and 
states that $V(\ell)$ saturates quite rapidly increasing $\ell$. 
This fact is closely related to the short-rangedness of the entanglement between 
different points of the system. In Refs.~\cite{tagliais} and~\cite{nostro} 
it was found instead a deviation from this law for long-range systems outside of criticality, 
resulting in a $V(\ell)$ increasing logarithmically with $\ell$, as for critical short-range systems. 

In this Section, $V(\ell)$ will be derived out of criticality for $\alpha <2$, 
using the correlation functions computed in Section \ref{corrET} by means of the effective theory. 
We refer to Ref.~\cite{ares} for an analytic computation of $V(\ell)$ in the long-range paired Kitaev 
chains based on the properties of Toeplitz matrices.

We focus at the beginning on the contribution to $V(\ell)$ from $S_{\mathrm{AN}}^{(\alpha<2)}$, $V^{(\alpha<2)}(\ell)$. 
This quantity can be calculated following Refs.~\cite{callan,casini,peschel} via the formula
\beq
V^{(\alpha<2)}(\ell) = - \, \mathrm{Tr} \, \Big((1-C) \, \log_2 (1-C) + C \,  \log_2 C \Big) 
\eeq
where
\begin{equation}
C(x, y) =  \langle 0|\psi_{\mathrm{H}}(0, x) \psi^{\dagger}_H(0,y) |0\rangle= 
\frac{1}{L}    \sum_p  \, e^{-i p (x-y)} \, \Bigg( 
\frac{M
 - \, \gamma_1   \, \mathrm{sign}(\mathrm{sin} \, p) \,  |\mathrm{sin} \, p|^{\beta} }{\sqrt{|\mathrm{sin} \, p|^{2\beta} + M^2}}  \,+ \,  \gamma_0   \Bigg) \, \gamma^0 
\label{propnto2}
\end{equation}
with $\{p\} = \frac{2 \pi}{L} \, n$, $n = 0, \dots, L-1$.
Eq.~\eqref{propnto2} is obtained through a
discretization of \eqref{propnto}.
The von Neumann entropy
$V^{(\alpha<2)}(\ell)$ is then fitted by the formula 
$a + \frac{c_\m{AN}^{\mathrm{(VNE)}}}{3} \, \log_2  \big( \frac{\ell}{L} \big)$, 
expected in critical short-range one-dimensional systems 
and introduced for long-range systems in Ref.~\cite{tagliais} also near criticality.  
 
We checked, as a first step, that the extracted values 
for $c_\m{AN}^{\mathrm{(VNE)}}$ converge quite rapidly with $L$ increasing, 
as discussed in Refs.~\cite{callan,casini,nez}. The results of our study are the following:
\begin{itemize} 

\item $c_\m{AN}^{\mathrm{(VNE)}} = \frac{1}{2}$ if $M = 0$. Indeed no $\beta$ dependence arises in 
Eq.~\eqref{propnto2} in this case, 
so that we recover the value for a massless Majorana theory, as for the critical short-range Ising model 
\cite{wilczek,calabrese} at $\beta  = 1$;

\item $c_\m{AN}^{\mathrm{(VNE)}} = 0$ if $M \to \infty$: for the same reason as above, 
no $\beta$ dependence occurs in Eq.~\eqref{propnto2} and $V(\ell)$ saturates, increasing $\ell$, to a constant 
value, as for gapped short-range systems;

\item if $M$ is finite, 
$c_\m{AN}^{\mathrm{(VNE)}}$ passes abruptly (for $L \to \infty$) from 
$c_\m{AN}^{\mathrm{(VNE)}} = \frac{1}{2}$ 
(as when $\alpha = 0$) to $c_\m{AN}^{\mathrm{(VNE)}} = 0$ 
(as when $\alpha \to \infty$). 
The change of value for $c_\m{AN}^{\mathrm{(VNE)}}$ arises 
at a certain critical $\alpha$ rapidly increasing with $M$. 

\end{itemize}

Recalling that $M \to \infty$ for $\alpha > 1$ and $M \to 0 $ for $\alpha < 1$ 
(as seen in Section \ref{ET}), we conclude that near criticality ${c_\m{AN}^{\mathrm{(VNE)}}}  = 0$ 
if $1< \alpha < 2$  and ${c_\m{AN}^{\mathrm{(VNE)}}}  = \frac{1}{2} $ if $\alpha < 1$. 

The contribution $V_\m{D}(\ell)$ to $V(\ell)$ from the Dirac action $S_{\mathrm{D}}$ is known \cite{calabrese} 
to scale as $a + \frac{c_{\mathrm{D}}^{(\mathrm{VNE})}}{3} \, \log_2  \big( \frac{\ell}{L} \big) $ with 
$c_{\mathrm{D}}^{(\mathrm{VNE})}=0$ in the massive regime.
Thus, we can conclude that the parameter $c^{\mathrm{(VNE)}}$, governing the violation of the area-law for 
$V(\ell)$ and obtained from the total action in Eq.~\eqref{aztot},  coincides in the massive regime and for all the $\alpha$
 with $c_\m{AN}^{\mathrm{(VNE)}}$. 

On the massless line $\mu=1$ and  for every $\alpha$, 
$c_{\mathrm{D}}^{(\mathrm{VNE})}=c=\frac{1}{2}$  ($c$ being  the central charge of the Ising conformal theory) \cite{wilczek,calabrese}. Then 
 $c^{\mathrm{(VNE)}}$  
shows  at $\alpha=1$ the similar discontinuity as $c_{\m{AN}}^{\mathrm{(VNE)}}$. In particular $c^{\mathrm{(VNE)}} = \frac{1}{2}$ for $\alpha >1$ 
and $c^{\mathrm{(VNE)}}= 1$ for $\alpha <1$. 
The contribution to $c^{\mathrm{(VNE)}}$  by $S_{\mathrm{D}}$ exactly at criticality and at $\alpha<2$ is justified  by the fact that  there  the leading behavior for the two-fermions correlations in the large $R$ limit still originates from the modes near  $k = 0$ (as well as near $k = \pi$ if $\alpha <1$, see Section \ref{latcorr}). 
 
In the light of the discussion in Section \ref{ET} about the evolution of the mass parameters along the RG, 
we are led to conclude that the discontinuity for $c^{\mathrm{(VNE)}}$ is due 
to the different behavior of $M$ across $\alpha = 1$, behaving the model Eq.~\eqref{Ham} 
close to criticality effectively as massless below this threshold. 
The same behavior for $c^{\mathrm{(VNE)}}$ has been 
exactly found in Ref.~\cite{ares}: this confirms the correctness of our calculation 
and, at more at the basic level, of our effective action (\ref{aztot}), having two different contributions.

In the range $\alpha > 2$ one gets $c_{\m{AN}}^{\mathrm{(VNE)}}=0$ near criticality and 
$c_{\m{AN}}^{\mathrm{(VNE)}}=1/2$ at criticality. 
A reliable estimation can be given retaining in Eq.~\eqref{IRlower2} only the terms $\propto p^2$ and $\propto p^{\alpha}$, seen in Section \ref{corrET} 
to assure a correct qualitative behavior for the static correlation functions out of criticality. 
This choice takes to a matrix correlation
\begin{equation}
C(x, y) =  \langle 0|\psi_{\mathrm{H}}(0, x) \psi^{\dagger}_H(0,y) |0\rangle= 
\frac{1}{L}    \sum_p  \, e^{-i p (x-y)} \, \Bigg[ 
\frac{M
 - \, \gamma_1   \, \mathrm{sign}(\mathrm{sin} \, p) \,  (|\mathrm{sin} \, p|^{\beta} + r(\beta) |\mathrm{sin}(p)|)}{\sqrt{(|\mathrm{sin} \, p|^{\beta} + r(\beta) |\mathrm{sin}(p)|)^2+ M^2 }} \,+ \,  \gamma_0   \Bigg] \, \gamma^0 \,
\label{propnto3}
\end{equation}
analogous to the one in Eq.~\eqref{propnto2} and with $r(\beta) = \frac{c_{1,0}}{a_0} \, \frac{1}{b^{\beta-1}} \gg 1$. 
Since in this range for $\alpha$ it holds $M \to \infty$ along the RG, as for $1<\alpha<2$, 
the same considerations done above leads to conclude that no area-law violation, logarithmic or more pronounced, 
arises out of criticality (while at criticality the same behavior as for the short-range Kitaev chain occurs). This conclusion matches the results in \cite{nostro,ares}.

However a deeper analysis is required for the von Neumann entropy at $\alpha> 2$  to investigate on 
possible sub-logarithmic deviations from the area-law, 
suggested by the algebraic decay tails (even if suppressed at large $\alpha$) 
for the static correlation functions and expected due to the long-range pairings in (\ref{Ham}). 
For this purpose an improved RG procedure is also necessary, to properly take 
into account the other RG subdominant terms. 

We finally notice that the discussion above shows the 
potential difficulty to obtain the correct behavior 
for the von Neumann entropy at every finite $\alpha$ by an effective action having a single contribution (see also the end of the Subsection \ref{discussion}),  
providing a check for the correctness of our action in Eq.~\eqref{aztot}.


\section{Critical line $\mu = -1$}
\label{muminus}

For the sake of completeness and comparison, in this Section we briefly 
discuss the properties of the model around the critical line $\mu=-1$. This line differs from the line
$\mu = 1$, since, as found in \cite{nostro}, the model at finite $\alpha$ is no longer symmetric 
under $\mu \to -\mu$.
An analysis similar as the one carried out for $\mu = 1$ can be 
provided now when $\alpha>1$ (for $\alpha<1$ the Hamiltonian in Eq.~\eqref{Ham} acquires a mass gap).

In these conditions the minimum of the lattice spectrum arises at
the edges of the Brillouin zone ($k = \pm \pi$). For this reason a unique term occurs in $H_L^{(b)}$ 
after the decimation step and consequently  in the renormalized effective action $S$. 
Moreover exactly at criticality, close to the minimum the energy grows linearly at  for $\alpha > 2$, while it scales as $k^{\alpha -1}$ for $\alpha<2$. 
These facts imply together that at criticality the breakdown of conformal symmetry appears at $\alpha  = 2$, 
where the spectrum ceases to be linear (there the quasiparticle velocity starts to diverge, 
see Appendix \ref{expansion}).
More in detail, if $\alpha <2$ 
the same calculation as the one performed around $\mu = 1$ leads again to an effective action as in Eq.~\eqref{lren3}. 
There notably  a mass $m \propto |\mu +1|$, as in $S_{\mathrm{D}}$, is present, vanishing at criticality  and diverging along the RG flow outside of it, while no anomalous mass  $M \propto |\mu-1|$ occurs.  
In this way both the exponential and algebraic parts visible in the correlation $g_{1}^{(\mathrm{lat})} (R)$
develop from $S$, conversely to the case $\mu = 1$. 
Moreover the critical theory is invariant under the asymmetric rescaling symmetry $x \to \lambda^{\frac{1}{\beta}} \, x$, $\tau \to \lambda \, \tau$, as for $\mu \approx 1$ and $\alpha<1$  (see Section \ref{ET}).

If $\alpha > 2$ 
the same effective action as the one around $\mu = 1$ is also found, 
with dispersive terms $\propto p^2$ and $\propto p^{\beta}$
(see Section \ref{corrET} and \ref{VNE}). Related to this fact, 
exactly at criticality the breakdown of the conformal symmetry is again signaled by the equality $\beta = 1$ between the power exponents of these terms. 
As for $\alpha<2$, the mass $m$ vanishes exactly at criticality and out of it diverges along the RG flow, as in the short-range Ising model. Moreover, since here it holds $\beta >1$, in this point a massless Dirac action $S_{\mathrm{D}}$ 
is recovered. Our results are summarized in Table~\ref{table2}.

\begin{table}
\begin{center}
\begin{tabular}{ |l | c |  c| }
  \hline
  $\mu=-1$           & Critical point     & Near criticality \\ \hline
  $\alpha>2$ & CI and ELI         & ELI          \\ \hline
  $1<\alpha<2$ & Non CI and Non ELI & Non ELI          \\ \hline
\end{tabular}
\caption{Summary of obtained results for the critical line $\mu=-1$: as in Table~\ref{table1}, CI stands 
for Conformal Invariance, and ELI for Effective Lorentz Invariance.}\label{table2}
\end{center}
\label{tabella1}
\end{table}

An analogous analysis as in Section \ref{latcorr} for the critical line $\mu = 1$ can be repeated 
in the present case. Again we find a good qualitative agreement at large $R$ between the lattice predictions 
and the ones from the effective theory theory described above.

In the light of the discussion in Section \ref{VNE},  the presence of a single term in the effective action 
and the behavior of $m$ both at and outside of criticality implies immediately that for $\alpha>1$ the area-law 
violation for the von Neumann entropy results in the values $c^{\mathrm{(VNE)}} = \frac{1}{2}$ 
at criticality and $c^{\mathrm{(VNE)}} = 0$ outside of it. This result is in agreement 
with the result derived in \cite{ares}.


\section{Critical exponents}
\label{crexp}

We described in the previous Sections the simultaneous presence of two contributions $S_{\mathrm{D}}$ and $S_{\mathrm{AN}}$
in the effective action (\ref{aztot}) out of criticality, as well as some related effects. 
A further notable consequence of this result
is that the long-range Kitaev chain is an example of a long-range model   
where the value of $\alpha$ at which the conformal symmetry is surely 
broken does not coincide with the value where the critical exponents 
of the corresponding short-range model  at $\alpha \to \infty$ change. 

To show this fact, we analyze on the lattice the order parameter \cite{refpar}
\beq
m_{\alpha} (\mu) = \lim_{R\to\infty}\sqrt{ \mathrm{det} \, G_{R,0} (\alpha , \mu)} \, ,
\eeq
where
\beq
 \mathrm{det} \, G_{R,0} (\alpha , \mu) = \mathrm{det} \Big[\delta_{R,0} + 2 \braket{a^\dag_R a^\dag_0} +2 \braket{a^\dag_R a_0}\Big].
\eeq
This parameter characterizes when 
$\alpha \to \infty$ the paramagnetic-(anti)ferromagnetic
quantum phase transition of the short-range Ising model. Indeed this chain
is mapped via Jordan-Wigner transformation to the  short-range Kitaev model.
In particular $m_{\infty}(\mu)$  
coincides \cite{muss} with the average magnetization
of the short-range Ising chain; 
it has non vanishing values only for $|\mu| <1$ and approaching $|\mu| = 1$ 
from below it scales as  $m_{\infty}(\mu) \propto (1- |\mu|)^{\beta}$, $\beta = \frac{1}{8}$.

{\em A priori} it is not fully obvious that $m_{\alpha} (\mu)$ is a good order parameter
even at finite $\alpha$, however we assume its correctness also in this range, motivated by the absence of 
discontinuities in $\alpha$ out of the critical lines, at least up to $\alpha =1$.
In favor of our assumption we further notice that, as in the short-range limit, $m_{\alpha} (\mu)$ is not  vanishing only
if $|\mu|<1$, tending to zero approaching the critical lines. Again we focus first 
on the critical line $\mu = 1$.

An explicit numerical evaluation on the lattice of the critical exponent $\beta$ with which $m_{\alpha} (\mu)$ 
vanishes approaching from below the line $\mu=1$ shows that, for each positive $\alpha$, $\beta$ is equal 
(within numerical precision) to $1/8$, seen above to be the critical exponent $\beta$ for 
the short-range limit.

This fact can be explained remembering 
that $\braket{a^\dag_i a^\dag_j}$ and $\braket{a^\dag_i a_j}$ take contributions both from the modes close to 
$k   = 0$ and to $k   = \pm \pi$ described in our effective formulation by $S_{\mathrm{D}}$ and 
$S_{\mathrm{AN}}$, respectively. When the sum of multilinear terms resulting from 
$\mathrm{det} \, G_{ij} (\alpha, \mu)$ is performed, these modes give rise also to isolated terms in the sum. 
For this reason it is straightforward to relate the dominant part in the scaling of $m_{\alpha}(\mu)$, equal as 
for the short-range Ising model, to the contributions by the modes around $k =0$: indeed they are leading for 
$m_{\alpha}(\mu)$ in the $R \to \infty$ limit, as it happens for the two fermions correlations exactly at criticality.

A similar result is obtained for the critical exponent $\alpha$ 
(in the standard definition of the critical exponents \cite{huang}, not to be confused with the decay exponent of 
the pairing in the Hamiltonian of Eq. (\ref{Ham})) related to the effective specific heat $c(\mu)$, defined as 
(see \cite{delfino04,um2006} and references therein)
\begin{equation}\label{eqn:SpecificHeat}
c(\mu) = \frac{\partial^2}{\partial  \mu ^2} \, \int_{- \pi}^{\pi} \mathrm{d} k \, \lambda_{\alpha}(k) = \int_{- \pi}^{\pi} \mathrm{d} k \, \frac{f_\alpha^2(k)}{\lambda^3_{\alpha}(k)}.
\end{equation}   
This quantity turns out to have the same logarithmic  divergence as the short-range Ising model for each positive 
decay exponent $\alpha$ of the pairing, 
as the integrand in Eq.~\eqref{eqn:SpecificHeat} displays a logarithmic singularity for $k \to 0$ when $\mu=1$.

Calling $\alpha^{\ast}$ the value of the decay exponent such that for $\alpha > \alpha^{\ast}$ all the 
critical exponents of the short-range model are retrieved, we would then conclude that for 
the critical line $\mu=1$ of our model $\alpha^{\ast}=0$.

We noticee that $\alpha^{\ast}$ does not coincide in this case  
with the value at which CS 
is explicitly broken (equal to $2$ for the considered case).

A more interesting scenario occurs for the critical exponents of Ising order parameter in the critical line 
$\mu=-1$. 
Indeed they are the same as the short-range Ising model if $\alpha >2$, while they change below this threshold.
In particular the critical exponent $\beta$ starts to decrease, while the effective specific heat 
$c(\mu)$ does not display the logarithmic divergence, because the integrand in Eq. \eqref{eqn:SpecificHeat} shows an integrable singularity $\propto 1/|k-\pi|^{\alpha-1}$
when $k\to\pi$ and $1<\alpha<2$, resulting in a continuous $c(\mu)$ as Fig.\ref{fig:SpecificHeat} shows. 
From Eq. \eqref{eqn:SpecificHeat} it is clear that the disappearance of this divergence originates
from the change of the spectrum below $\alpha = 2$, the same inducing the breakdown of CS. 
Similarly, preliminary computations show that also the magnetic critical 
exponent $\beta$ changes below $\alpha = 2$.
These facts highlight the deep relation between the loss of CS and
the change of the critical exponents in the present example.

The different behavior of the exponents in the present case compared to the case $\mu \approx 1$ appears 
to be compatible with the presence now of an unique term in the effective action, 
being not a Dirac action if $\alpha<2$. 
\begin{figure}
\includegraphics[scale=0.4]{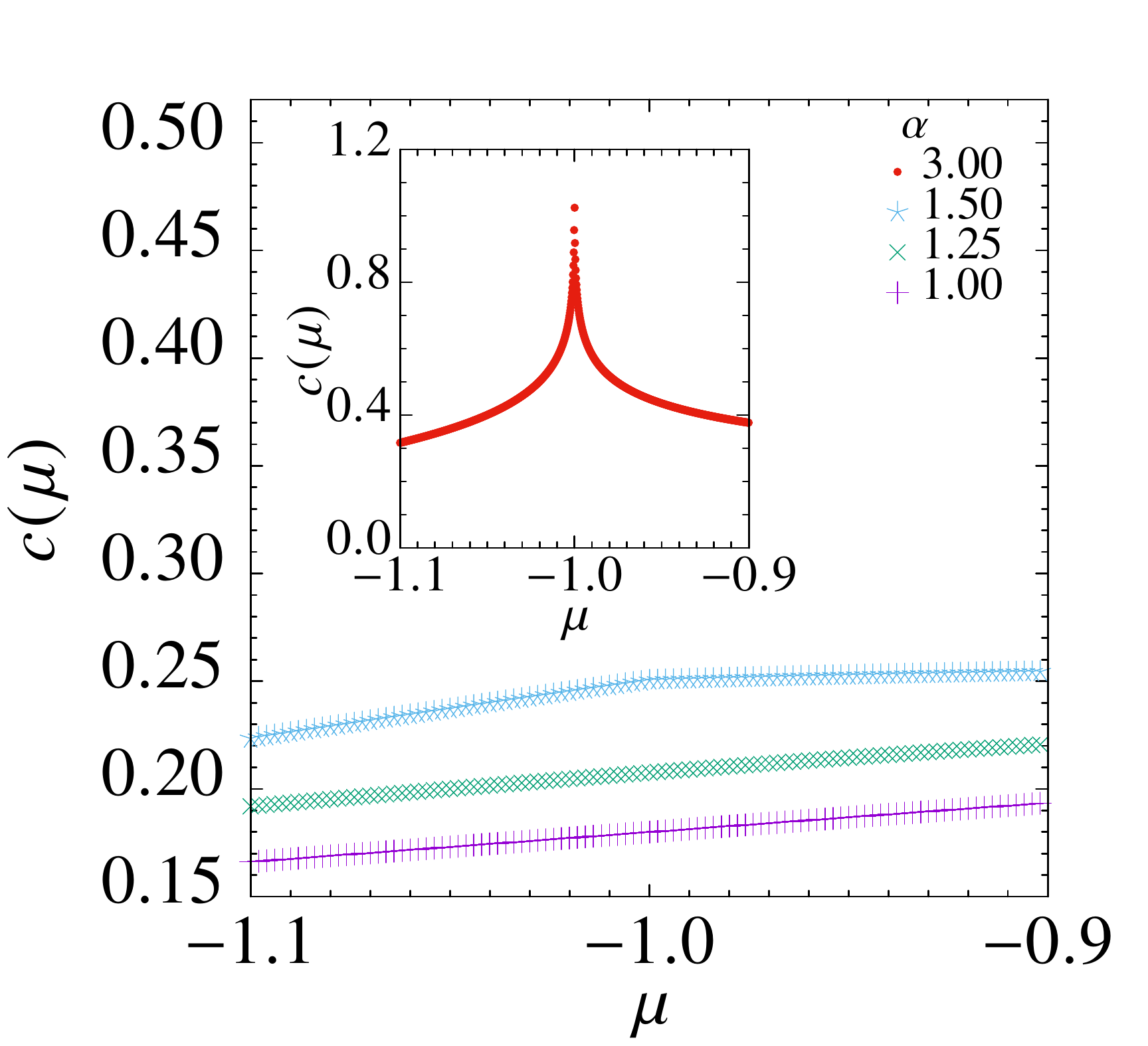}[h!]
\caption{Main panel: The effective specific heat $c(\mu)$ near the critical line $\mu=-1$ for $1<\alpha<2$ is continuous as the integrand in Eq. \eqref{eqn:SpecificHeat} shows an integrable  divergence. For comparison the inset shows the logarithmic divergence of $c(\mu)$ for $\alpha>2$.}
\label{fig:SpecificHeat}
\end{figure}


\section{Conclusions and Perspectives} 
\label{concl}

Motivated by the interest in investigating the symmetries of critical points on long-range systems 
and specifically whether they possess (or not) conformal symmetry (CS), 
we determined and studied the effective theories close and at the critical points of the Kitaev chain with long-range 
pairing decaying by a power-law of the distance. The main reason for the choice of this model 
relies on the fact that it is a quadratic solvable model, displaying non-trivial quantum phase transitions and 
critical behavior. 
 
Using a renormalization group approach, we derived the effective theory of the long-range 
paired Kitaev chain close to criticality, showing that the effective theory around the critical line $\mu = 1$ 
is the sum of two commuting terms: a Dirac
action plus a conformal symmetry breaking term. 
For $\alpha<2$ the effective theory is dominated 
by the latter term, implying the explicit breakdown of the conformal symmetry at 
the critical point. 
This breakdown is found to be caused by the contribution -- usually neglected in the determination of 
the effective theory for short-range critical models -- of the modes at the edges of the Brillouin zone. 
Indeed, even if these points are not minima of the energy spectrum, the modes around them do 
not decouple from the critical dynamics, determining an anomalous contribution to the effective theory, 
responsible for the conformal symmetry breakdown. 
 We expect that this scenario can be qualitatively extended 
to other long-range models displaying quasiparticle excitations in ballistic regime. 
For instance, the hybrid static correlation functions
of the long-range Ising model have been discussed to have, 
at least in regimes describable satisfactorily by a spin wave approach, a double contribution 
from different sets of modes \cite{cirac05}, exactly as for the long-range Kitaev chains studied here and in \cite{paperdouble}.

Our results are based on the identification of the contribution 
of the high-energy modes at the edges of the Brillouin zone, which are explicitly taken into account into the 
lower energy part of the renormalized action. A remarkable advantage of using both the terms in the action $S$ is then
that both near and at criticality all the leading contributions of the correlations are reproduced in a
natural way. 
In order to test the validity of the obtained effective theory, we compared two fermions connected correlations 
calculated from it with the corresponding lattice results, 
finding a good agreement. 
In particular for both of them we derived an 
exponential-plus-algebraic decay of correlation functions at every finite $\alpha$ with the decay rates obtained from the two schemes that are coinciding. It is found that at intermediate and large length scales 
this behavior results from the interplay of the modes with minimum energy 
(responsible of the exponential decay) with the ones at the edges of the Brillouin zone 
(giving the power-law tail), 
the same responsible of the explicit conformal symmetry breakdown  
for small $\alpha$. 

The effective theory also allows us to highlight a direct
relation between the observed hybrid decay of correlations 
and the breakdown of the 
Lorentz invariance emerging close to criticality. 
The links between effective Lorentz invariance and conformal symmetry have been analyzed.

Finally, using the effective theory  we computed the violation of the area-law for 
the von Neumann entropy out of criticality and in the regime 
of small $\alpha$, finding agreement with previous analytical and numerical results.
Importantly, this agreement also confirms the correctness of our effective action (\ref{aztot}), having two different contributions.

A notable consequence of our results 
is that the long-range Kitaev chain on the critical line $\mu = 1$ is an example of a long-range model   
where the value of $\alpha$ at which the conformal symmetry is surely 
broken does not coincide with the value where the critical exponents 
of the corresponding short-range model at $\alpha \to \infty$ change. 
Indeed an explicit numerical evaluation on the lattice of 
the critical exponent $\beta$ with which 
the Ising order parameter 
vanishes approaching from below the critical line $\mu =1$ shows that, for each positive value of $\alpha$, $\beta$ is equal 
(within numerical precision) to $1/8$, which is the value of the critical exponent $\beta$ for 
the short-range Ising model in a transverse field,  in turn directly mapped via Jordan-Wigner transformation to the $\alpha \to \infty$ short-range Kitaev model.
A similar result is obtained for the specific heat critical exponent $\alpha$ 
\big(in the standard definition of the critical exponents \cite{huang}, not to be confused with the decay exponent of the pairing in the Hamiltonian of Eq. (\ref{Ham})\big), 
which turns out to have the same logarithmic  divergence of the short-range Ising model for each positive decay 
exponent of the pairing. 
Calling $\alpha^{\ast}$ the value of the decay exponent such that for $\alpha > \alpha^{\ast}$ 
all the critical exponents 
of the short-range model are retrieved, we would then conclude that for our model $\alpha^{\ast}=0$. If from one side 
we may ascribe this behavior to the quadratic free 
nature of the long-range paired Kitaev model, 
from the other side it is enough to show 
that $\alpha^{\ast}$ does not coincide in general with the value at which conformal symmetry 
is explicitly broken (equal to $2$ for the considered case).

At variance, for the critical line $\mu = -1$ the Ising critical exponents change where 
the conformal symmetry is broken. This behavior occurs due to 
the presence of an unique term in the effective action, being a Dirac action for $\alpha>2$ and 
not a Dirac action for $\alpha<2$. In turn the appearance of this unique 
term derives from the fact that the minimum of the energy is indeed at the edges of the Brillouin zone. 

As future work, we mention that a first point 
concerns the justification of the anomalous quasiparticles weights assumed along the RG flow.
Indeed this assumption has been made mainly motivated by the hybrid decay behavior of the static correlation 
functions on the lattice. From the mathematical point of view, the non-decoupling during 
the decimation procedure of the high-energy modes at the edges of the  Brillouin zone
looks to be related with the requirement of smoothness of the RG flow, encoded in the infinite differentiability 
of the regularized Hamiltonian $H_L(b)$. Indeed at every finite $\alpha$, $H_L(b)$ is differentiable only a finite number of times, since the same property holds for $\lambda_{\alpha} (k)$. 
The same singularities were shown responsible for the algebraic decay of the static correlations at large separations \cite{nostro} and for the emergence of the area-law logarithmic violation for von Neumann entropy \cite{ares}. 
Motivating and clarifying 
this conjecture certainly deserves future effort.

Another important subject of further work is to understand the 
role and the implications of the RG irrelevant terms in the anomalous part of the effective theory, also 
for $\alpha > 2$. Indeed, in spite of their behavior along the RG flow, 
they are responsible for the power-law decay tails of the static correlation functions, also observed on the lattice. 
For the same reason these terms are expected to determine sub-logarithmic
deviations from the area-law for the von Neumann entropy.  On the contrary, RG subleading terms do not affect the phase diagram of the model in Eq. (\ref{Ham}), in particular the appearance of new phase(s) at $\alpha <1$. 
A complete clarification of these issues would require an exact RG computation, without {\em a priori} 
(even though physically motivated) choices of the quasiparticle weights along the RG flow.

We finally comment that an interesting line of research starting from the results of the present paper consists 
in a systematic comparison of the symmetries of the long-range paired 
Kitaev phases/critical points with the corresponding ones for the long-range Ising models, 
where an explicit interaction between the excitations is involved, 
and in extending the results and the techniques discussed here
to  higher dimensional long-range models.\\

{\bf Acknowledgements --} 
The authors thank in a special way J. Polonyi for his remarks on 
the different hierarchy of the quasiparticles weight along RG. They 
also acknowledge very useful discussions with  
N.~Defenu, A.~Gorshkov, M.~Maghrebi, and M.~Mannarelli. L.~L. thanks  L.~Salasnich for support.
A.~T. gratefully acknowledges discussions in IIP (Natal) with 
T.~Apollaro, F.~Ares and A.~Queiroz.  
L.~L., G.~P., and D.~V. acknowledge support by the ERC-St Grant ColdSIM 
(Grant No. 307688). A.~T. acknowledges support from the
Italian PRIN "Fenomeni quantistici collettivi: dai sistemi fortemente correlati ai simulatori quantistici" 
(PRIN 2010\textunderscore2010LLKJBX).\\

{\bf Note --} During the final part of this work, useful correspondence with S. Rychkov 
informed us about a very recent paper \cite{slava} on the conformal symmetry of the 
classical long-range Ising model: a systematic comparison of long-range Kitaev and Ising models 
is required in order to understand the apparently different occurrence, 
as a function of $\alpha$, of the conformal symmetry at their critical points. We also mention 
the arXiv submission \cite{Magh_bis}, in which the continuous symmetry breaking in low-dimensional quantum 
systems with long-range interactions is studied. Finally, very recently it also appeared a paper 
dealing with global quenches on the Hamiltonian (\ref{Ham}) and with the consequent spread of the mutual information \cite{wouters2015}.

\appendix 

\section{Properties of the excitation spectrum}
\label{expansion}

In this Appendix we analyze the behavior of $\lambda_\alpha(k)$ and of
its first $k$-derivative (quasiparticle velocity) around $k = \pm \pi$. For this purpose
we exploit the periodicity of the Brillouin zone $k = k + 2 \pi$, 
allowing to restrict ourselves to the limit $k \to \pi^{\pm}$. 

Exploiting the series expansion of the polylogarithms \cite{nist}, we find that for $p = k -\pi \to 0^{\pm}$
and for non integer $\alpha$: 
\begin{equation}
\label{eqn:TaylorOfEff}
f_{\alpha}(p+ \pi)  =2 \cos{\frac{\pi \alpha}{2}} \, \Gamma(1-\alpha) \, 
\mathrm{sign}(p) |p|^{\alpha-1} + 2 \sum_{n=1}^{\infty}  \, \Big(\sin{\frac{\pi n}{2}}\Big) \, 
\frac{\zeta(\alpha-n)}{n!} \, p^{n} \equiv a_{0} \, p^{\alpha-1}+\sum_{n=1}^{\infty} c_n \, p^n,
\end{equation}
and
\begin{equation}
f^2_{\alpha}(p + \pi) = 4 \cos^2{\frac{\pi \alpha}{2}}\,\Gamma^2(1-\alpha) \, p^{2\alpha-2} +
8 \cos\frac{\pi \alpha}{2} \zeta(\alpha-1) \, \Gamma(1-\alpha) \, p^\alpha   + 4 \, \zeta^2(\alpha-1) \, p^2 
+ { \dots \, ,} 
\label{expans2}
\end{equation}
where the symbol $\dots$ indicates the terms not inducing the leading contribution of $\mathrm{Im} \, G_{\alpha} (i \, p)$ in Eq.~\eqref{corr+2}. In this way, when $p\to 0^{\pm}$ one has 
\begin{equation}
\lambda^2_\alpha (p+ \pi) = (\mu+1)^2 + 4 \cos^2\left(\frac{\pi\alpha}{2}\right) \Gamma^2(1-\alpha) \, p^{2\alpha-2} +8 \cos\frac{\pi \alpha}{2} \zeta(\alpha-1)\Gamma(1-\alpha) \, p^\alpha - \left((\mu+1)- 4 \zeta^2(\alpha-1) \right) p^2. 
\label{exp}
\end{equation}
If $\alpha <2$ the first two terms in Eq.~\eqref{exp} are dominating for $p \to 0^{\pm}$, 
while the last one is for $\alpha > 2$.

The previous formulas allow to determine the lattice quasiparticle velocity for $p\to 0^{\pm}$:
\begin{equation}
\frac{\de \lambda_\alpha(p + \pi)}{\de p} = \frac{1}{\lambda_\alpha(p + \pi)} \left(-\sin (p + \pi) (\cos (p + \pi) + \mu ) + f_{\alpha}(p +\pi) \frac{\de f_\alpha(p + \pi)}{\de p}\right).
\end{equation}
We focus in particular on the case $p \to 0^{+}$. 
From Eq.~\eqref{eqn:TaylorOfEff} we have on the critical line $\mu=1$:
\begin{equation}
\left.\frac{\de \lambda_\alpha( p + \pi)}{\de p}\right|_{p \to 0} \sim \begin{cases}
p^{\alpha-2}  \to \infty   	& \text{for }  \alpha<1 \, ; \\
p^{2\alpha-3} \to \infty 	& \text{for } 1<\alpha<3/2 \, ; \\
p^{2\alpha-3} \to 0		 	& \text{for } 3/2<\alpha<2 \, ; \\
p 			  \to 0 		& \text{for }  \alpha>2 \, ,
\end{cases}.
\label{vel}
\end{equation}
showing a divergence if $\alpha<\frac{3}{2}$. The situation is different
along the critical line $\mu=-1$. There we have in particular the following cases:
\begin{equation}
 \left.\frac{\de \lambda_\alpha( p + \pi)}{\de p}\right|_{p \to 0} \sim 
\begin{cases}
p^{\alpha-2} \to \infty & \text{ for } \alpha<2 \, ; \\
 \mathrm{constant} & \text{ for } \alpha>2 \, ,
\end{cases}
\end{equation}
showing a divergence for $\alpha<2$.


\section{Fields and correlation functions}
\label{soluzioni}

The tight-binding matrix Hamiltonian corresponding to Eq.~\eqref{Ham} in the main text is
{
\begin{equation} h_\alpha(k) = 
\begin{pmatrix} 
 -(\mu -2 \omega \, \cos{k})   \, \, \, \, i \Delta \, f_{\alpha}(k + \pi) \\
- i \Delta \, f_{\alpha}(k+ \pi) \,\, \, \, \, \, (\mu - 2 \omega \, \cos{k} )\\
   \end{pmatrix}
\end{equation}
}
in the basis 
$\begin{pmatrix} a_k \\
   a_{-k}^{\dagger} 
\end{pmatrix}
$. 

In the limit $k \to \pi$ and setting $p =  (k - \pi)$ we obtain the 
Hamiltonian $h_{\mathrm{AN}}(p)= \gamma_0 \, \Big(\gamma_1 \, p^{\beta} + M \Big)$, with 
$\gamma_0 = -\sigma_3$ and $\gamma_1 = -i \, \sigma_1$, $p^{\beta}$ being defined as in Section \ref{ET} of the main text.

The eigenfunctions of $h_\alpha(k)$ are 
$$u_{+} (k) = \begin{pmatrix} \cos{\theta_k} \\ i \, \sin{\theta_k}  \end{pmatrix}$$ and 
$$u_{-} (-k) = \begin{pmatrix} -i \, \sin{\theta_k} \\   \cos{\theta_k}  \end{pmatrix}$$
(respectively with positive and negative energy; $k >0$ here), 
where {$\sin^2{\theta_k} = \frac{1}{2} \, \Big(1 + \frac{\mu -\cos{k} }{\lambda_\alpha(k)} \Big)$ and 
$\cos^2{\theta_k} = \frac{1}{2} \, \Big(1 - \frac{\mu -\cos{k}}{\lambda_\alpha(k)} \Big)$}
are the Bogoliubov coefficients. The ground state is defined as in the main text at the end of Section 
\ref{model}. These formulas follow directly from the expression for 
the Bogoliubov transformation:
\begin{equation}
\begin{pmatrix}a_{k}\\  a^\dag_{-k} \end{pmatrix}= V^{\dagger} \begin{pmatrix} \eta_{k} \\  \eta_{-k}^\dag\end{pmatrix}
\label{bog}
\end{equation}
with
\begin{equation}
V =
\begin{pmatrix}
\cos{\theta_k} &  i \sin{\theta_k}  \\ 
- i \sin{\theta_k} & \cos{\theta_k}  \end{pmatrix}.
\end{equation}
The $\eta_k$ label the annihilation operators for the Bogoliubov quasiparticles.

The eigenfunctions of $ h_{\mathrm{AN}}(p)$, $\tilde{u}_{\pm} (p)$, 
have the same functional form as the lattice ones $u_{\pm} (\pm k)$
but with $\lambda_\alpha(p) = \sqrt{(p^{\beta})^2 + M^2 }$.
Using these expressions we obtain for the Majorana field $\psi_{\mathrm{H}}(x,t)$ \cite{pal}:
\beq
\psi_{\mathrm{H}}(t,x) =  \int  \,\frac{\mathrm{d} p}{2 \pi} \, \frac{1}{\sqrt{2 \lambda(p)}} \, \Bigg( \tilde{u} (p) \, e^{-i
\big(E(k) t - p x\big)} \, b(p) \,+ \,  \gamma_0 \, C \, \tilde{u}^{*}(p) \, e^{i
\big(E(k) t - p  x\big)} \, b^{\dagger}(p) \Bigg). 
\label{quant}
\eeq   
The positive and negative-energy solutions $\tilde{u}_{\pm} (\pm p)$ and the related operators $\{\eta_p, \eta^{\dagger}_{-p}\}$ 
are collectively denoted by $\tilde{u}(p)$ and $b(p) $, while $C = \sigma_2$ is the charge conjugation matrix. We notice that 
the normalization of Eq.~\eqref{quant}, adopted in the calculation of Eq.~\eqref{propnto}, differs from the one induced by the field rescaling in Eq.~\eqref{H2}
\beq
\psi_{\mathrm{H}}(t,x) =  \int  \, \mathrm{d} p \, \frac{1}{\sqrt{2 \pi L}} \, \Bigg( \tilde{u} (p) \, e^{-i
\big(E(k) t - p x\big)} \, b(p) \,+ \,  \gamma_0 \, C \, \tilde{u}^{*}(p) \, e^{i
\big(E(k) t - p  x\big)} \, b^{\dagger}(p) \Bigg) 	, 
\label{quant2}
\eeq   
and also used in literature (see e.g. \cite{sak}).
However the two points correlation functions do not depend on the normalization assumed.

The field $\psi_{\mathrm{L}}(t,x)$ quantizes in the same manner as Eq.~\eqref{quant}, as well as the field constructed by 
the lattice solutions $h(k)$.

From the structure of the solutions of $h(k)$ and  $ h_{\mathrm{AN}}(p)$, as well as of the fields constructed by them, 
it is possible to infer that the (matrix) propagator in real space 
$$D(x^{\mu}-y^{\mu}) = \langle 0|\psi_{\mathrm{H}}(0, x) \bar{\psi}_{\mathrm{H}}(0,y) |0\rangle$$ 
corresponds to the matrix of the correlation functions on the lattice
\beq
\tilde{C}(i, j) =  \frac{1}{L}    \sum_k  \, e^{-i k (i-j)} \, 
\begin{pmatrix} 
\mathrm{cos}^2   \theta_k   \, \, \, \, \,  -i \,\mathrm{sin}  \theta_k \, \mathrm{cos}  \theta_k\\
i \, \mathrm{sin}   \theta_k \mathrm{cos} \,  \theta_k \, \, \, \, \, \mathrm{sin}^2   \theta_k\\
   \end{pmatrix} = \begin{pmatrix} -\langle c_i c^{\dagger}_j \rangle \, \, \, \langle c^{\dagger}_i c^{\dagger}_j \rangle  \\ 
{} \, \, \, \, {}\\
- \langle c_i c_j \rangle \, \, \,  \langle c^{\dagger}_i \, c_j \rangle 
\end{pmatrix}
\label{corrC}
\eeq
 (up to a negligible $\delta_{x,0}$ function present in the diagonal entries of (\ref{corrC})), with $k=- \pi + 2\pi \left(n +  1/2\right)/L$ 
and $0 \leq n< L$.


\section{Breakdown of conformal symmetry for $S_{\mathrm{AN}}$}
\label{breaking_CS}

In this Section we provide details on the 
the behavior of the Euclidean action Eq.~\eqref{lren3} 
under the global conformal transformations.
Their infinitesimal forms read:
$$
x^{\mu} \to x^{\prime \, \mu} = x^{\mu} + \epsilon^{\mu},
$$
with $ x^{\mu} \equiv (\tau,x)$ and {\em i)} $
\epsilon^{\mu} = a^{\mu}$  (translations); {\em ii)} 
$\epsilon^{\mu} = \omega^{\mu \nu } \, x_{\nu}$ (rotations); 
{\em iii)} $\epsilon^{\mu} = \lambda \, x^{\mu}$ (dilatations); 
{\em iv)} $\epsilon^{\mu} = b^{\mu} \, x^2 -2 x^{\mu} b \cdot x$ (special conformal transformations).
 $a^{\mu}$, $\lambda$, and $b^{\mu}$ are constant parameters and $ b_{\mu} x^{\mu} \equiv b \cdot x$. 
One can check straightforwardly that Eq.~\eqref{lren3} is invariant under translations.

Under dilatations $(x, \tau) \to \lambda \, (x, \tau)$ 
the action Eq.~\eqref{lren3} becomes:
\beq
S_{\mathrm{AN}}^{(\alpha<2)}  =    
\int  \frac{1}{\lambda^2} \, \m{d} x  \, \m{d}\tau \, \bar{\psi}_{\mathrm{H}} (x,\tau) \, \lambda^{c_{\psi}} \,  \Big(\lambda \, \gamma_0 \, \partial_\tau   \,  + \,  \lambda^{\beta} \, \gamma_1 \,    \partial_x^{\beta}  \Big) \,  \psi_{\mathrm{H}} (x,\tau) \, \lambda^{c_{\psi}}.
\label{lagr2}
\eeq
We see that, independently from the value of $c_{\psi}$, 
a global reabsorption of the $\lambda$ factors is impossible if $\beta \neq 1$, then Eq.~\eqref{lagr2} is not invariant 
under dilatations. 
However, as discussed in the Section \ref{csbreak} of the main text, a non isotropic scale invariance remains,  defined 
as $x \to \lambda^{\frac{1}{\beta}} \, x$, $\tau \to \lambda \, \tau$. These 
transformations do not belong to the conformal group. 

In order to better understand the behavior under rotations, 
it is instructive to consider at the beginning the massless Dirac 
case $\beta = 1$: 
\beq
S_{\mathrm{AN}}^{(\alpha<2;  \beta=1)}  = 
\int  \m{d} x  \, \m{d}\tau \, 
\bar{\psi}_{\mathrm{H}} (x^{\mu}) \,  \gamma_{\mu} \, \partial^{\mu}   \,  \psi_{\mathrm{H}} (x^{\mu}).
\label{lagr3}
\eeq
In this case the invariance under rotations follows directly from the well known Lorentz invariance of the 
Minkowskian action corresponding to Eq.~\eqref{lagr3} \cite{peskin}. In the
demonstration of the Lorentz invariance it is crucial that the product $ \gamma_{\mu} \, \partial^{\mu} $
is a scalar, resulting from the contraction of two bi-vectors. 
For this reason we infer immediately that if $\beta \neq 1$ no rotational symmetry survives 
for Eq.~\eqref{lagr3}, since the $\tau$ and $x$ derivatives cannot be organized in a bi-vector transforming covariantly.  

The discussion of the invariance under special conformal transformations (SCT) requires a particular care.
To perform it properly we have to recover first the defining properties of the conformal group.
A conformal transformation is defined \cite{dif} as an invertible mapping $(\tau, x) \to (\tau^{\prime}, x^{\prime})$ that leaves invariant the metric tensor
$g_{\mu \nu} (\tau , x)$, up to a local scale factor $\Lambda (\tau , x)$:
\beq
g_{\mu \nu} (\tau , x) \to g^{\prime}_{\mu \nu} (\tau^{\prime} , x^{\prime}) = \Lambda (\tau , x) \, g_{\mu \nu} (\tau , x) 	\,,
\label{SCT}
\eeq
where $\Lambda (\tau , x)   \equiv J(\tau, x)^{\frac{d}{2}}$, $d = 2$ being the
dimension of the Euclidean space $(\tau, x)$ and $J(\tau , x) $ the Jacobian of the transformation. For translation and rotations $J(\tau , x) = 1$, while for dilatations $J(\tau,x) = \lambda^ D$ (in the notation of the present paper), then under such transformations the metric tensor $g_{\mu \nu} (\tau , x)$ stays constant or it undergoes just a global rescaling. 

The situation is completely different for the SCT, since now $J(\tau , x)= \frac{1}{1- 2 \, b^{\rho} x_{\rho} +  b_{\delta} b^{\delta} \,  x_{\eta} x^{\eta}}$, being $b^{\rho}$
the bi-vector of the parameters characterizing a SCT. Because of the $J(\tau, x)$ associated to them, the SCT induce a non trivial (local)  rescaling of the metric tensor $g_{\mu \nu} (\tau , x)$. In particular, even if before a transformation $g_{\mu \nu} (\tau , x) = \eta_{\mu \nu}$ ($\eta_{\mu \nu}= \mathrm{diag(1, 1)}$), after it in general $g^{\prime}_{\mu \nu} (\tau^{\prime} , x^{\prime})$ is not diagonal any longer. 
For this reason one is forced to analyze the invariance under SCT in a general \emph{curved} Euclidean space-time. 
Again we discuss first the Dirac action $\beta = 1$, that is written in such a space-time as \big($(\tau,x) \equiv x$ for 
the sake of brevity\big):
\beq
S_{\mathrm{AN} ; \mathrm{CURV}}^{(\alpha<2;  \beta=1)}  = 
\int  \m{d} x  \, 
\sqrt{\mathrm{det} \, g^{\mu \nu} (x)} \, \, \, \psi^{\dagger}_{\mathrm{H}} (x) \tilde{\gamma}^0 (x)\,  \tilde{\gamma}^{\mu}(x) \, g_{\mu \nu} (x) \, D^{\nu} (x)  \,  \psi_{\mathrm{H}} (x), 
\label{lagrcurvnov}
\eeq
with $\{\tilde{\gamma}^{\mu}(x) , \tilde{\gamma}^{\nu}(x)\} = 2 \, g^{\mu\nu}(x)$ and
$D^{\nu} (\tau, x)$ being the covariant derivative \cite{birrell}. 
Notice that we wrote $S_{\mathrm{AN} ; \mathrm{CURV}}^{(\alpha<2;  \beta=1)} $ without the use of the vielbeins coordinates, as most widely done in literature (see e.g. \cite{birrell}), since our choice makes easier the following discussion.
The transformation law of  $\psi_{\mathrm{H}} (x)$ under SCT reads \cite{dif,birrell}: 
\beq
\psi_{\mathrm{H}} (x) \to \Lambda(x)^{\frac{d_{\psi}}{D}}  \, \psi_{\mathrm{H}} (x)  
\eeq
($d_{\psi} = \frac{1}{2}$ as in the main text); the same transformation holds for $D^{\nu} (x)  \,  \psi_{\mathrm{H}} (x)$ by the fundamental property of the covariant derivative. 
Since $g_{\mu \alpha} \, g^{\alpha \nu} = \delta^{\nu}_{\mu}$ we have also that $\tilde{\gamma}^{\mu}(x) \to \Lambda(x)^{-\frac{1}{2}} \, \tilde{\gamma}^{\mu}(x)$
and   $\sqrt{\mathrm{det} \, g^{\mu \nu} (x)} \to \Lambda(x)^{-\frac{D}{2}} \, \sqrt{\mathrm{det} \, g^{\mu \nu} (x)}$ (so that in particular $ \m{d} x  \, 
\sqrt{\mathrm{det} \, g^{\mu \nu} (x)}$ is a scalar, as well known \cite{landau2}). 
These rescaling properties, together with the requirement of invariance for the fermionic anti-commutation relations, imply that the factors $\Lambda(x)^{l}$ ($l$ in the set described above) appearing in the action Eq.~\eqref{lagrcurvnov} after the transformations Eq.~\eqref{SCT} cancel each other. For this reason
Eq.~\eqref{lagrcurvnov} is invariant under SCT.
Analyzing in detail the steps of the demonstration above, it results that the invariance of Eq.~\eqref{lagrcurvnov} under SCT depends crucially on the fact that the covariant derivative  $D^{\nu} (x)  \,  \psi_{\mathrm{H}} (x)$ transforms 
as $\psi_{\mathrm{H}} (x)$. Is it clear that this property does not hold any longer if $\beta \neq 1$ (the concept of covariant derivative itself being questionable in this case), implying directly that   Eq.~\eqref{lagrcurvnov}
 is not any longer invariant under SCT if $\beta \neq 1$.


\section{Scaling of the density of the ground-state energy}
\label{scaling_GS}

We discuss here the conformal symmetry breakdown in connection to the scaling 
with $L$ 
of the ground-state energy density on the lattice $e_0(\alpha, L)$, also in relation 
with the analysis in \cite{nostro}, 
where it was concluded that conformal symmetry breaking occurs below $\alpha = \frac{3}{2}$. 

The lattice energy density of the ground state in the limit $L \to \infty$ reads
$$e_0 (\alpha, \infty) = \frac{1}{2 \pi}  \int_{-\pi}^{\pi} \, \m{d} k \, 
\lambda_{\alpha}(k)$$
(notice that it is finite in the same limit,  
then no Kac rescaling \cite{libro} is needed).

For a $(1+1)$D conformal theory the textbook formula for $e_0$ is 
given by \cite{dif,muss}: 
\beq
e_0(\alpha, L) = e_0(\alpha, \infty) - \pi \, v_F \, c/(6 L^2),
\label{CFTscal}
\eeq
where $c$ labels the central charge of the conformal theory and 
$v_F$ is the Fermi velocity at the conformal point \cite{dif,muss}.
 
Using the Euler-MacLaurin formula for the polylogarithms
\cite{grad} in the $L \to \infty $ limit, one obtains at $\mu = 1$ 
\cite{nostro}
\beq
e_0(\alpha, L) = e_0(\alpha, \infty) +  \pi \, \left[\lambda_\alpha'(0) - \lambda_\alpha'\left(\pi \right)\right]/(12 L^2),
\eeq 
where $\lambda_\alpha'(k)$ labels the first derivative { of leading term in the expansion} of 
$\lambda_\alpha(k)$ around $k = \pm \pi$ (see details in Appendix 
\ref{expansion}). This approximation amounts to discard the subleading terms in the effective action above $\alpha = 2$.
 
While the finite contribution from $\lambda_\alpha'\left(0 \right)$ 
allows to recover the critical Ising theory 
with $c = \frac{1}{2}$ above $\alpha = \frac{3}{2}$, 
the divergence of  $\lambda_\alpha'\left(\pi \right)$ at 
$\alpha < \frac{3}{2}$ was discussed to be 
a possible signature of the conformal symmetry breakdown \cite{nostro}. 
However the Euler-MacLaurin expansion has 
higher-order terms in powers of $\frac{1}{L}$:
\begin{equation}
e_0(\alpha, L) = e_0(\alpha, \infty) + \pi \left[\lambda_\alpha'(0) - 
\lambda_\alpha'\left(\pi \right)\right]/(12 L^2) + 
\pi^3 \left[\lambda_\alpha'''(0) - 
\lambda_\alpha'''\left(\pi \right)\right]/(720 L^4) + o\big(1/L^4\big).
\label{viol}
\end{equation}
While the quantity $\lambda_\alpha'''\left(0 \right)$ vanishes, 
$\lambda_\alpha'''\left(\pi \right)$ is divergent if $\alpha <2$, 
signaling the breakdown of conformal symmetry also for $\frac{3}{2} < \alpha <2$. 
Notice that, since $\lambda_\alpha'\left(\pm \pi \right) \propto k$ at 
$\alpha>2$ (see Appendix 
\ref{expansion}), no further breakdown it is found at higher orders 
for the expansion Eq.~\eqref{viol}.

This result can be numerically tested 
by fitting $e_0(\alpha, L)$, drawn for different lattice sizes $L$, 
by the conformal scaling law Eq.~\eqref{CFTscal}. 
It is found that this law works very well for $\alpha >2$, 
yielding the Ising conformal charge $c = \frac{1}{2}$ up to numerical errors. 
At variance, the fit with Eq.~\eqref{CFTscal}  works very poorly below $\alpha=2$, 
and a not well defined $c$, oscillating or even negative, is obtained, 
probably as an effect of the divergent terms in Eq.~\eqref{viol}.


\section{Correlations of the long-range Kitaev chain}
\label{appcorr}

In this Appendix we provide details on the analytical computation of the two-points 
correlations $g_1^{\text{(lat)}}(R) = \langle a^\dag_R a_0 \rangle$ and $g^{\text{(lat; anom)}}_1(R) = \langle 
a^\dag_R a^\dag_0 \rangle$ \cite{paperdouble}. 
As the model is quadratic, all the correlation functions can be built 
from $g_1^{\text{(lat)}}(R)$ and $g^{\text{(lat; anom)}}_1(R)$ by mean of the Wick's theorem. \\
The Brillouin zone 
is chosen to be $[0, 2 \pi]$, moreover we focus on the range $\mu >0$.

In the limit $L\to\infty$, 
$g_1^{\text{(lat)}}(R)$ becomes 
\begin{equation}
g_1^{\text{(lat)}}(R) = - \frac{1}{2\pi} \int_{0}^{2\pi} \m{d} k \, e^{i k R } \, \mathcal{G}_\alpha(k),
\label{eqn:NormalCorrelatorContinuum} 
\end{equation}
with
\begin{equation}\label{eqn:IntegrandGreenFunction}
\mathcal{G}_\alpha(k)=\frac{\cos{k}+\mu}{2\lambda_\alpha(k)}.
\end{equation}
To evaluate the integral in Eq.~\eqref{eqn:NormalCorrelatorContinuum}, we deform the linear path $[0, 2 \pi]$ to the integration contour in Fig.~\ref{fig:contornoHybrid} 
and use the Cauchy theorem:
\begin{equation}
g_1^{\text{(lat)}}(R)  = -\frac{1}{2\pi}\lim_{M\to \infty }\left( \int_{\mathcal{C}_0}  + \int_{\mathcal{L}_{-}} + \int_{\mathcal{L}_{+}} + \int_{\mathcal{C}_{2\pi}}  \right) \m{d} z \, e^{i z R} \,\mathcal{G}_\alpha(z)
\label{intot}
\end{equation}
with $z=k+i y$ and $M$ defined as in Fig.~\ref{fig:contornoHybrid}. 
We can neglect the contributions from \(\mathcal{C}_\perp\) and \(\mathcal{C}'_\perp\), as they vanish when 
$M \to \infty$.

\begin{figure}[h!!]\centering
\includegraphics[scale=0.85]{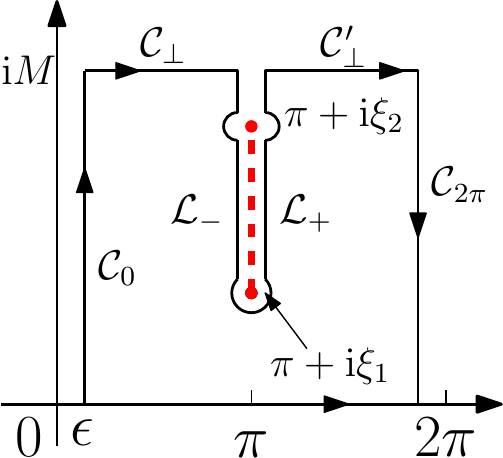}
\caption{Deformed integration contour to evaluate the integral 
in Eq.~\eqref{eqn:NormalCorrelatorContinuum}. The dashed line is the branch cut of the square root in the denominator of the integrand.} \label{fig:contornoHybrid}
\end{figure}

The contours $\mathcal{L}_\pm$ are chosen since the square root in denominator of 
Eq.~\eqref{eqn:IntegrandGreenFunction} displays two complex roots $\pi + i \, \xi_{1,2}$  given by 
the solutions of the equation
\begin{equation}\label{eqn:correlationLength}
(\mu -\cosh \xi_{1,2})^2+ f_\alpha(\pi + i \xi_{1,2})^2 =0.
\end{equation}
A branch cut arises between the roots above, as shown in Fig.~\ref{fig:contornoHybrid}.

Let us now analyze the different contributions to Eq.~\eqref{intot} from the various parts of the path 
in Fig.~\ref{fig:contornoHybrid}. We will conclude in the next following that the parts \(\mathcal{L}_{\pm}\), 
involving momenta close to $k = \pi$, give the exponential decaying part of $g_1^{\text{(lat)}}(R)$, 
while $\mathcal{C}_{0}$ and $\mathcal{C}_{2\pi}$, involving momenta close to $k = 0$, 
give the power-law decaying part. 

Let us consider separately these two contributions:

\subsubsection{Exponential part}

If $\xi_1 < y < \xi_2$ we have $\mathcal{G}_\alpha(\pi^+ + i y) = \mathcal{G}^*_\alpha(\pi^-  + i y)$ 
due to the branch cut of the square root in Eq.~\eqref{eqn:IntegrandGreenFunction}. 
The integrals on the two lines \(\mathcal{L}_{-}\) and  \(\mathcal{L}_{+}\) give
\begin{equation}
I_{\mathcal{L}_{-}} +I_{\mathcal{L}_{+}}  = \frac{e^{i \pi R} e^{-\xi_1 R}}{\pi} \, \int_{0}^{\xi_2}  \m{d} y \, 
e^{- y R} \, \mathrm{Im} \mathcal{G}_\alpha(\pi^{+}+i (y+\xi_1)).
\label{int1}
\end{equation}
The previous integral is a Laplace-type integral \cite{Ablowitz2003}. 
We can get its leading behavior, first by replacing $\xi_2$ with infinity, 
as the difference is exponentially suppressed, and then  by integrating the $y\to 0$ limit in 
$\mathcal{G}_\alpha(\pi^{+}+i (y+\xi_1))$.
This part reads:
\begin{equation}
\mathcal{G}_\alpha(\pi^{+}+i (y+\xi_1)) \sim  \frac{A_\alpha(\mu)}{i \sqrt{y}}\quad \text{ if } y\to 0
\end{equation}
with 
\begin{equation}\label{eqn:CoefficientExponentialPart}
A_\alpha(\mu)=\frac{\mu-\cosh\xi_1}
{2\sqrt{2} 
\Big[\mathrm{Li}_{\alpha -1}(-e^{\xi_1})+\mathrm{Li}_{\alpha -1}(-e^{- \xi_1)})\Big]^{1/2}
\Big[\mathrm{Li}_{\alpha}(-e^{\xi_1})-\mathrm{Li}_{\alpha}(-e^{- \xi_1)})\Big]^{1/2}},
\end{equation}
and $\xi_{1}$ is the smallest solution of
\begin{equation}
(\mu - \cosh \xi_{1,2})^2- \left(\mathrm{Li}_{\alpha}(-e^{-\xi_{1,2}}) - 
\mathrm{Li}_{\alpha}(-e^{\xi_{1,2}})\right)^2=0.
\end{equation}
In this way the integration in Eq.~\eqref{int1} yields 
the exponentially decaying part of $g_1^{\text{(lat)}}(R)$:
\begin{equation}
I_{\mathcal{L}_{-}}+I_{\mathcal{L}_{+}} = A_\alpha(\mu) \frac{e^{i \pi R}}{\sqrt{\pi}}\frac{e^{-\xi_1 R}}{\sqrt{R}}.
\end{equation}

\subsubsection{Power-law part}

On $\mathcal{C}_{0}$ it is 
\begin{equation}
I_{\mathcal{C}_0} =-\frac{1}{2\pi} \int_{\mathcal{C}_0} e^{i z R} \,\mathcal{G}_\alpha(z) \m{d} z 
= -\frac{i}{2\pi} \int_{0}^{\infty} e^{-y  R} \, \mathcal{G}_\alpha(\epsilon+i y) \m{d} y,
\end{equation}
while on $\mathcal{C}_{2\pi}$ (with $z=2\pi - \epsilon + i y$) 
\begin{equation}
I_{\mathcal{C}_{2\pi}} =-\frac{1}{2\pi} \int_{\mathcal{C}_2\pi} e^{i z R} \,\mathcal{G}_\alpha(z) \m{d} z 
= \frac{i}{2\pi} \int_{0}^{\infty} e^{-y  R}\, \mathcal{G}_\alpha(2\pi-\epsilon+i y) \m{d} y.
\end{equation}

The integrals on $\mathcal{C}_0$ and $\mathcal{C}_{2\pi}$ sum to
\begin{equation}
I_{\mathcal{C}_{0}}+ I_{\mathcal{C}_{2\pi}} = \frac{1}{\pi}  \int_0^\infty  \m{d} y \,e^{- y R} \, \mathrm{Im}  
(\mathcal{G}_{\alpha}(i  y)).
\label{eqn:IntegralZeroTwoPi}
\end{equation}
Again we can evaluate the asymptotic behavior of Eq.~\eqref{eqn:IntegralZeroTwoPi} by computing the $y\to 0$ part 
of $\mathcal{G}_{\alpha}(i  y)$ and 
then integrating it \cite{Ablowitz2003}. The series expansion of the polylogarithm  \cite{nist}
$$
\mathrm{Li}_{\alpha}(e^{\pm y}) = - \Gamma(1-\alpha)(\mp y)^{\alpha-1}+ \sum_{j=0}^\infty \frac{\zeta(\alpha-j)}{j!} (\pm y)^j
$$
fixes the main contribution to the imaginary part of $\mathcal{G}_\alpha(i y)$ for $y \to 0$, coming from the expression:
\begin{equation}\label{eqn:GFunctionZeroMomentum}
\mathcal{G}_\alpha(i y) \sim \frac{\mu+1}{2\sqrt{(\mu+1)^2-\Gamma^2(1-\alpha)(e^{i \pi \alpha}+1)^2 y^{2\alpha-2}+4 \Gamma(1-\alpha)(e^{i \pi \alpha}+1)\zeta(\alpha-1) \, y^\alpha}}.
\end{equation}
In Eq.~\eqref{eqn:GFunctionZeroMomentum} we discarded  higher integer $y$-powers, 
not contributing to  $\mathrm{Im} \, \mathcal{G}_\alpha(i y)$, as well as the higher non integer powers, whose contribution to $\mathrm{Im} \, \mathcal{G}_\alpha(i y)$ is suppressed in  the limit $y \to 0$.

Putting Eq.~\eqref{eqn:GFunctionZeroMomentum} in Eq.~\eqref{eqn:IntegralZeroTwoPi} yields:
\begin{equation}
I_0 + I_{2\pi}   = \begin{cases}
-\dfrac{ 2\alpha \zeta(\alpha-1)}{\signum{(\mu+1)}(1+\mu)^2}
\dfrac{1}{R^{\alpha+1}}  & \text{for $\alpha>2$}, \\[2em]
\dfrac{ \sin (\pi  \alpha ) \cos ^2\left(\frac{\pi  \alpha }{2}\right) \Gamma(2\alpha-1)} {\pi\signum{(\mu+1)}(\mu+1)^2} \dfrac{1}{R^{2\alpha-1}}  &  \text{for $1<\alpha<2$}, \\[2em]
\dfrac{(\mu+1)(1-\alpha)}{4\pi} \dfrac{1}{R^{2-\alpha}} & \text{for $\alpha<1$}. 
\end{cases}
\end{equation}

By collecting all the contributions, one obtains
\begin{equation}\label{eqn:TwoContributions}
\langle a^\dag_R a_0 \rangle = \frac{A_\alpha(\mu) e^{i \pi R}}{\sqrt{\pi}}\frac{e^{-\xi_1 R}}{\sqrt{R}} 
-\frac{2 \alpha \zeta(\alpha-1)}{\mathrm{sign}{(\mu+1)}(1+\mu)^2}\frac{1}{R^{\alpha+1}} \,\,\, \text{for } \alpha>2,
\end{equation}
\begin{equation}\label{eqn:TwoContributionsTwo}
\langle a^\dag_R a_0 \rangle =  \frac{A_\alpha(\mu)e^{i \pi R}}{\sqrt{\pi}}\frac{e^{-\xi_1 R}}{\sqrt{R}}  
+ \frac{\sin{(\pi  \alpha)} \cos^2\left(\frac{\pi  \alpha }{2}\right)} 
{\pi\mathrm{sign}{(\mu+1)}(\mu+1)^2} \frac{\Gamma(2\alpha-1)}{R^{2\alpha-1}} \,\,\, \text{for } 1<\alpha<2,
\end{equation}
and
\begin{equation}\label{eqn:TwoContributionsThree}
\langle a^\dag_R a_0 \rangle =  \frac{A_\alpha(\mu) e^{i \pi R}}{\sqrt{\pi}}\frac{e^{-\xi_1 R}}{\sqrt{R}} 
+ \frac{(\mu+1)(1-\alpha)}{4\pi} \frac{1}{R^{2-\alpha}} \,\,\, \text{for } \alpha<1.
\end{equation}

When $\alpha<1$ one can show that, even if Eq.~\eqref{eqn:correlationLength} has one or two solutions, 
the exponential part in Eq.~\eqref{eqn:TwoContributionsThree} is negligible with respect to the power-law tail. 
This gives effectively a pure power-law correlation function. 

\subsubsection{Anomalous correlation function}

For completeness we report here details of the  calculation for the
anomalous correlation function $g^{\text{(lat; anom)}}_1(R)=\langle a^\dag_R a^\dag_0 \rangle$, which is given by:
\begin{equation}\label{eqn:AnomalousCorrelatorContinuum}
g^{\text{(lat; anom)}}_1(R) = \frac{1}{2\pi}\int_{0}^{2\pi} \m{d} k \, e^{i k R} \, \mathcal{F}_\alpha(k) 
\end{equation}
with
\begin{equation}
\mathcal{F}_\alpha(k) = i \, \frac{f_\alpha(k)}{2\lambda_\alpha(k)}.
\end{equation}
Using the same integration contour as for $g^{\text{(lat)}}_1 (R)$, we get:
\begin{equation}
g^{\text{(lat; anom)}}_1 (R) =\frac{e^{i \pi R} e^{-\xi_1 R}}{\pi} \int_0^\infty \m{d} y \, e^{-y R} 
\mathcal{F}_\alpha(\pi^{+} + i (y+\xi_1))- \frac{1}{\pi} \int_0^\infty \m{d} y \, 
e^{-y R} \, \mathrm{Im} \mathcal{F}_\alpha(i y), 
\label{previous}
\end{equation}
showing both the exponential and the power-law contributions. In Eq.~\eqref{previous} $\xi_{1}$ 
is again the smallest solution of  Eq.~\eqref{eqn:correlationLength}. One obtains finally
\begin{equation}\label{eqn:AnomalousTotalUno}
g^{\text{(lat; anom)}}_1 (R) =
\frac{B_\alpha(\mu) e^{i \pi R}}{\sqrt{\pi}}\frac{e^{-\xi_1 R}}{\sqrt{R}} -\frac{1}{2 
|\mu+1|} \frac{1}{R^\alpha} \,\,\,\ \text{for } \alpha>1
\end{equation}
and 
\begin{equation}
g^{\text{(lat; anom)}}_1 (R) = \,\, \frac{B_\alpha(\mu) e^{i \pi R}}{\sqrt{\pi}}\frac{e^{-\xi_1 R}}{\sqrt{R}} -\frac{1}{2\pi}\frac{1}{R} 
\,\,\, \text{for } \alpha<1,
\end{equation}
with 
\begin{equation}\label{eqn:CoefficientExponentialPartAnomalous}
B_\alpha(\mu)=\frac{[\mathrm{Li}_{\alpha}(-e^{\xi_1})-\mathrm{Li}_{\alpha}(-e^{-\xi_1})]^{1/2}}{2\sqrt{2}
[\mathrm{Li}_{\alpha -1}
(-e^{\xi_1})+\mathrm{Li}_{\alpha -1}(-e^{- \xi_1)})]^{1/2}}.
\end{equation}


\section{Alternative computation of the long-range behavior of $\langle a^\dag_R a_0 \rangle$ 
by the effective theory (\ref{lren3})}
\label{altra_der}

We derive in this Appendix the asymptotic decay tail of the correlation function $\langle a^\dag_R a_0 \rangle$ 
obtained by the effective theory (\ref{lren3}) close to criticality. 
As written in the main text, Section \ref{corrET}, the asymptotic part of $\langle a^\dag_R a_0 \rangle$ and is given by
\beq
- \frac{1}{2 \pi}  \int  \m{d} p  \, \frac{M}{\sqrt{(p^{\beta} + r(\beta) \, p)^2 + M^2}} \, e^{i p R}  \equiv  
- \frac{1}{2 \pi}  \int  \m{d} p \, I\big(p, M\big) \,e^{i k R},
\label{primo}
\eeq
with $\beta = \alpha-1$. This integral can be evaluated along similar lines as the ones described in 
the previous Appendix for the lattice correlation functions, yielding the power-law asymptotical behaviors 
written in the main text.

A simpler approximated calculation can be as well performed, expanding $I\big(p, M\big)$ 
in powers of $\frac{1}{M}$ for $1< \alpha <2$ and of $M$ for $1< \alpha $ and integrating them in (\ref{primo}). 
In principle these expansions may lead to wrong results for Eq.~\eqref{primo}, due to the integration on momenta with large 
modulus, where the correctness of the expansions can cease to hold. 
However, since we are interested in the $R \to \infty$ limit, 
the momenta giving the dominant contribution to the integral (\ref{primo}) are the ones close to $p = 0$, then we expect the expansions of  
$I\big(p, M\big)$ to yield correct asymptotical results for the same integral.

In the case $1< \alpha < 2$ we obtain $I(p, M \to \infty) \approx 1- \frac{|p|^{2 \beta}}{2 \, 
M^2}$ and performing the integral Eq.~\eqref{primo} with this expansion we get an asymptotic decay tail  $\propto \frac{1}{R^{2 \alpha -1}}$, in agreement with Section \ref{corrET}. 
In the expansion we used the fact that our definition for $p^{\beta}$ on the real axis (see Section \ref{ET} in the main text) 
implies $|p^{\beta}| = |p|^{\beta}$. We observe that the divergence of $M$ for $1< \alpha < 2$ 
is essential to obtain the correct leading behavior for Eq.~\eqref{primo} in this regime.

For $\alpha <1$ we obtain $I\big(p, M \to 0\big) \approx \frac{M}{|p|^\beta}- \frac{M^2}
{2 |p|^{3 \beta}}$; performing again the integral in Eq.~\eqref{primo} on the leading term of this expansion for $R \to \infty $ 
we get a asymptotic decay tail  $\propto \frac{1}{R^{2- \alpha}}$, again in agreement with Section \ref{corrET}.

Notably the same expansion technique applied to the case $\beta  =1$ yields a decay $\propto \frac{1}{R^0}$, 
in agreement with the known exact result for Eq.~\eqref{primo}, predicting exponential decay. 


\section{Exponentially decaying terms from the integral in Eq.~\eqref{DOMINANT}}
\label{app:exp}

We show in this Appendix the presence of exponentially decaying terms from the integral in Eq.~\eqref{DOMINANT}. 
We specialize for simplicity our discussion to the case $\alpha = 1.75$ \big($\beta = \frac{3}{4}$\big). 
In this case, exploiting the definition for $p^{\beta}$ in Section \ref{ET}, 
the integral in Eq.~\eqref{DOMINANT} becomes:
\begin{equation}
I(R) =\mathrm{Re} \int_0^\infty \frac{e^{i p R}}{\sqrt{p^{3/2}+1}} \de p.
\label{intin}
\end{equation}
For sake of simplicity in Eq.~\eqref{intin} we set $M \equiv 1$; 
in each formula obtained below the dependence on of $M$  can be restored 
by the substitution $R \to R \, M^{\frac{1}{\beta}}$. 
The integral $I(R)$ is evaluated now by standard methods of complex analysis; 
notice also that from now on $z^{\beta}$ can be assumed without loss of generality to have the usual meaning: 
$z^{a} = |z|^{a} \, e^{i a \phi}$.

\begin{figure}[h!!]
\centering
\includegraphics[scale=0.8]{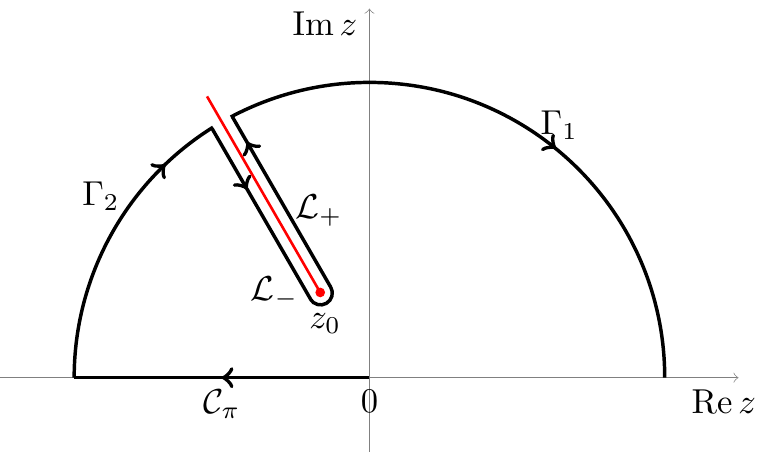}
\caption{Deformed integration contour to evaluate the integral in Eq. \eqref{intin}.}
\label{path1}
\end{figure}

The function $\sqrt{z^{3/2}+1}$ of the complex variable $z$ has a branch cut from $z_0=e^{2i\pi/3}$ to $\infty$, 
then the integral can be decomposed on the path in Fig.~\ref{path1} into the sum
\begin{equation}
\int_0^\infty \frac{e^{i p R}}{\sqrt{p^{3/2}+1}} \de p = I(\mathcal{C}_\pi) + I(\mathcal{L}_+) + I(\mathcal{L}_-)
\end{equation}
with
\begin{equation}
I(\mathcal{C}_\pi)=\int_{\mathcal{C}_\pi} \de z  \frac{e^{i z R}}{\sqrt{z^{3/2}+1}}  = -\int_0^\infty \de p \frac{e^{-i p R}}{\sqrt{e^{3 i \pi/2}p^{3/2}+1}},
\end{equation}
\begin{equation}
I(\mathcal{L}_+)=\int_{\mathcal{L}_+} \de z  \frac{e^{i z R}}{\sqrt{z^{3/2}+1}}  = e^{2 i\pi /3} \int_1^\infty \de \rho \frac{\exp[i e^{2 i\pi /3}\rho R]}{i\sqrt{\rho^{3/2}-1} }, 
\end{equation}
and
\begin{equation}
I(\mathcal{L}_-)=\int_{\mathcal{L}_-} \de z  \frac{e^{i z R}}{\sqrt{z^{3/2}+1}}  = e^{2 i\pi /3} \int_\infty^1 \de \rho \frac{\exp[i e^{2 i\pi /3}\rho R]}{-i\sqrt{\rho^{3/2}-1} } =e^{2 i\pi /3} \int_1^\infty \de \rho \frac{\exp[i e^{2 i\pi /3}\rho R]}{i\sqrt{\rho^{3/2}-1} } = I(\mathcal{L}_+).
\label{gammaeq}
\end{equation}
In particular the function  $\sqrt{z^{3/2}+1}$ satisfies for $ \rho>1$:
\begin{equation}
\sqrt{z^{3/2}+1} =
\begin{dcases}
i\sqrt{\rho^{3/2}-1} & z = \rho \, e^{i (2\pi /3 + \epsilon)}   \\
-i\sqrt{\rho^{3/2}-1} & z = \rho \, e^{i (2\pi /3 - \epsilon)}
\end{dcases}
\end{equation}
on the two different paths $\mathcal{L}_\pm$.

The contributions from $\mathcal{L}_\pm$ sum up to an exponential decaying term:
\begin{equation}
I(\mathcal{L}_+)+I(\mathcal{L}_-)  = B(R) \exp[-R \sin(2\pi /3)]  \, ,
\label{expon1}
\end{equation}
where
$$B(R) \equiv \frac{2}{i}e^{2\pi i /3}e^{-i R \cos(2\pi /3)} 
\int_0^\infty \de y \frac{\exp[-R y \sin{2 \pi /3} ]\exp[-R y i \cos{2 \pi /3} ]}{\sqrt{(y+1)^{3/2}-1} }$$

The integral $I(\mathcal{C}_\pi)$ has both an exponential 
and a power-law contribution. Indeed, as the function $\sqrt{e^{3 i \pi/2}z^{3/2}+1}$ 
has a branch from $z_1=e^{-i \pi/3}$ to $\infty$, by exploiting the path in Fig.~\ref{path2} we can write 
\begin{equation}
I(\mathcal{C}_\pi) =  -\int_0^\infty \de p \frac{e^{-i p R}}{\sqrt{e^{3 i \pi/2}p^{3/2}+1}} = 
i  \int_0^\infty \de y \frac{e^{-y R}}{\sqrt{e^{3 i \pi/2}(- i y)^{3/2}+1}} + 
2 \int_{\gamma_+} \de z  \frac{e^{- i z R}}{\sqrt{e^{3 i \pi/2}z^{3/2}+1}},
\end{equation}
where we used the fact that, as in Eq.~\eqref{gammaeq}, $I(\gamma_+) = I(\gamma_-)$.

\begin{figure}[h!!]
\centering
\includegraphics[scale=0.8]{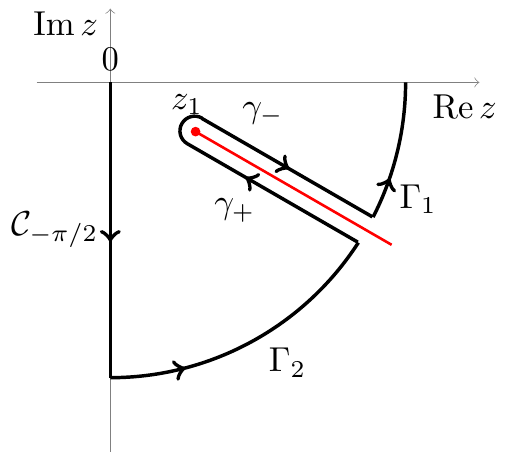}
\caption{Deformed integration contour to evaluate the integral in $I(\mathcal{C}_\pi)$.}
\label{path2}
\end{figure}
On $\gamma_+$, where $z=\rho \, e^{-i \pi/3}$, 
the function $\sqrt{e^{3 i \pi/2}z^{3/2}+1}$ becomes $i \sqrt{1-\rho^{3/2}}$ and $I(\gamma_+)$ is exponentially decaying:
\begin{equation}
\begin{split}
I(\gamma_+) & = \int_{\gamma_+} \de z  \frac{e^{- i z R}}{\sqrt{e^{3 i \pi/2}z^{3/2}+1}} 
=  e^{-i \pi/3}	\int_1^\infty \de \rho \frac{\exp[i e^{-i \pi/3} \rho R]}{i \sqrt{1-\rho^{3/2}}} = \\&
 = e^{-i \pi/3} \exp[-i \cos(\pi/3)  R - \sin(\pi/3)  R]	\int_0^\infty \de \rho \, \frac{\exp[e^{-i \pi/3} \rho R]}{i \sqrt{1-(\rho+1)^{3/2}}} \equiv \\& 
\equiv A(R) \exp[- \sin(\pi/3)  R].
\end{split}
\label{expon2}
\end{equation}

The power-law contributions are included instead in the term
\begin{equation}
\mathrm{Re} \int_0^\infty  \de y \frac{ i e^{-y R}}{\sqrt{e^{3 i \pi/2}(- i y)^{3/2}+1}}  \sim  \frac{1}{2} 
\cos\frac{\pi }{4} \int_0^\infty \de y \, e^{-y R} y^{3/2} = \cos\frac{\pi }{4} \frac{3  \sqrt{\pi }}{8 R^{5/2}}.
\label{centrale}
\end{equation}

Summing up all the contributions we obtain finally for the integral $I(R)$:
\begin{equation}
\begin{split}
I(R) & = \mathrm{Re} \int_0^\infty \frac{e^{i p R}}{\sqrt{p^{3/2}+1}} \de p =\\ &
=   \frac{3  \sqrt{\pi }\cos(\pi/4)}{8 R^{5/2}}
+ e^{-\sin(\pi/3)  R} \, \,  \mathrm{Re} \, A(R) + e^{-\sin(2\pi/3)  R} \, \, \mathrm{Re} \,  B(R),
\end{split}
\end{equation}
with the $A(R)$ and $B(R)$  from Eqs.~\eqref{expon1} and \eqref{expon2} decaying algebraically for $R\to\infty$.
These contributions are not present in the lattice results of Appendix \ref{appcorr}, since there only the exponential and algebraic leading contributions for $R \to \infty$ are derived.

When $1<\alpha < \frac{3}{2}$ the branch cut from $z_0$ moves out of the first and second quadrant and 
does not any longer contribute 
to the integral $I(R)$, by $I(\mathcal{L}_+)$. On the contrary,  the part $I(\mathcal{C}_\pi)$ is still present. However the exponential contribution $I(\gamma_{\pm})$ from it also disappears, since the corresponding branch cut moves out of the third quadrant. 
At the end it only remains:
\beq
I(R) = \mathrm{Re} \int_0^\infty  \de y \frac{ i e^{-y R}}{\sqrt{e^{i 2 \beta \pi}(- i y)^{2 \beta}+1}} ,
\eeq
as in Eq.~\eqref{centrale}. This integral gives substantially only algebraic decaying contributions. 
Possible exponentially decaying terms 
can occur only at very small $R$ and they are negligible.

If $\alpha <1$ the integral to be considered is 
\begin{equation}
I(R) =\mathrm{Re} \int_0^\infty \frac{p^{|\beta|}e^{i p R}}{\sqrt{p^{2|\beta|}+1}} \de p,
\label{intin3}
\end{equation}
from the second term of Eq.~\eqref{propnto}. The calculations proceed as for $\alpha >1$: similarly, 
for $1/2<\alpha<1$ exponential contributions does not any longer contribute 
to the integral $I(R)$.

We finally observe that, apart from exponential contributions, one has $I(R) \propto 1/R^{2\alpha-1}$ 
for $\alpha>1$ and $I(R) \propto 1/R^{2-\alpha}$ 
for $\alpha<1$.


\begin{thebibliography}{XX}

\bibitem{libro}
A. Campa, T. Dauxois, D. Fanelli, and S. Ruffo, 
{\em Physics of long-range interacting systems} (Oxford University Press, 2014).

\bibitem{cirac05}
X.-L. Deng, D. Porras, and J. I. Cirac, Phys. Rev. A {\bf 72}, 063407 (2005).

\bibitem{tagliais}
T. Koffel, M. Lewenstein, and L. Tagliacozzo, Phys. Rev. Lett. 
{\bf 109}, 267203 (2012). 
 
\bibitem{nostro}
D. Vodola, L. Lepori, E. Ercolessi, A. V. Gorshkov, and
G. Pupillo, Phys. Rev. Lett. {\bf 113}, 156402 (2014).

\bibitem{paperdouble}
D. Vodola, L. Lepori, E. Ercolessi, and G. Pupillo, 
New J. Phys. \textbf{18} 015001 (2016).

\bibitem{hast}
M. B. Hastings and T. Koma, Comm. Math. Phys. {\bf 265}, 781 (2006).

\bibitem{hauke2013}
P. Hauke and L. Tagliacozzo, Phys. Rev. Lett. {\bf 111}, 207202 (2013). 

\bibitem{eisert2014}
J. Eisert, M. van den Worm, S. R. Manmana, and M. Kastner, 
Phys. Rev. Lett. {\bf 111}, 260401 (2013).
 
\bibitem{metivier2014}
D. Metivier, R. Bachelard, and M. Kastner,
Phys. Rev. Lett. {\bf 112}, 210601 (2014). 

\bibitem{noinf} 
Z.-X. Gong,  M. Foss-Feig,  S. Michalakis, and A. V. Gorshkov,
Phys. Rev. Lett. {\bf 113}, 030602 (2014).
 
\bibitem{damanik2014}
D. Damanik, M. Lukic, W. Yessen, and M. Lemm, 
Phys. Rev. Lett. \textbf{113}, 127202 (2014).
  
\bibitem{nbound} 
M. Foss-Feig, Z.-X. Gong, C. W. Clark, and A. V. Gorshkov, 
Phys. Rev. Lett. {\bf 114}, 157201 (2015).

\bibitem{storch2015}
D. M. Storch, M. van den Worm, and M. Kastner, 
New J. Phys. {\bf17}, 063021 (2015).

\bibitem{carleo}
L. Cevolani, G. Carleo, and L. Sanchez-Palencia, 
Phys. Rev. A {\bf 92}, 041603(R) (2015). 
  
\bibitem{kastner2015ent}
M. Kastner, New J. Phys. \textbf{17}, 123024 (2015).

\bibitem{kuwahara2015}
T. Kuwahara, New J. Phys. 18 (2016) 053034.

\bibitem{growth}
J. Schachenmayer, B. P. Lanyon, C. F. Roos, and
A. J. Daley, Phys. Rev. X {\bf 3}, 031015 (2013).

\bibitem{santos2015}
L. F. Santos, F. Borgonovi, G. L. Celardo, Phys. Rev. Lett. 116 (2016), 250402.

\bibitem{Childress2006}
L. Childress, M. V. Gurudev Dutt, J. M. Taylor, A. S. Zibrov, F. Jelezko, J. Wrachtrup, P. R. Hemmer, and M. D. Lukin, Science {\bf 314}, 281 (2006).

\bibitem{Balasubramanian2009}
G. Balasubramanian, P. Neumann, D. Twitchen, M. Markham, R. Kolesov, N. Mizuochi, J. Isoya, J. Achard, J. Beck, J. Tissler, et al., Nature Mater. {\bf 8}, 383 (2009).

\bibitem{Weber2010}
J. R. Weber, W. F. Koehl, J. B. Varley, A. Janotti, B. B. Buckley, C. G. Van de Walle, 
and D. D. Awschalom, Proc. Natl. Acad. Sci. {\bf 107}, 8513 (2010).

\bibitem{tech0} S. Gopalakrishnan, B. L. Lev, and P. M. Goldbart, 
Phys. Rev. Lett. {\bf 107}, 277201 (2011). 

\bibitem{tech1} J. W. Britton, B. C. Sawyer, A. C. Keith, C.-C. J. Wang, 
J. K. Freericks, H. Uys, M. J. Biercuk, and J. J. Bollinger, Nature 
{\bf 484}, 489 (2012).

\bibitem{Schauss2012}
P. Schau\ss, M. Cheneau, M. Endres, T. Fukuhara, S. Hild, A. Omran, T. Pohl, C. Gross, S. Kuhr, and I. Bloch, Nature 
{\bf 491}, 87 (2012).

\bibitem{Aikawa2012}
K. Aikawa, A. Frisch, M. Mark, S. Baier, A. Rietzler, R. Grimm, and F. Ferlaino, Phys. Rev. Lett. {\bf 108}, 210401 (2012).

\bibitem{Lu2012}
M. Lu, N. Q. Burdick, and B. L. Lev, Phys. Rev. Lett. {\bf 108}, 215301 (2012).

\bibitem{Firstenberg2013}
O. Firstenberg, T. Peyronel, Q.-Y. Liang, A. V. Gor- shkov, M. D. Lukin, and V. Vuletic, Nature 
{\bf 502}, 71 (2013).

\bibitem{Yan2013}
B. Yan, S. A. Moses, B. Gadway, J. P. Covey, K. R. A. Hazzard, A. M. Rey, D. S. Jin, and J. Ye, Nature 
{\bf 501}, 521 (2013).

\bibitem{Dolde2013}
F. Dolde, I. Jakobi, B. Naydenov, N. Zhao, S. Pezzagna, C. Trautmann, J. Meijer, P. Neumann, F. Jelezko, and J. Wrachtrup, Nature Phys. {\bf 9}, 139 (2013).

\bibitem{Islam2013}
R. Islam, C. Senko, W. C. Campbell, S. Korenblit, J. Smith, A. Lee, E. E. Edwards, C.-C. J. Wang, J. K. Freericks, and C. Monroe, Science {\bf 340}, 583 (2013).

\bibitem{exp1}
P. Richerme, Z.-X. Gong, A. Lee, C. Senko, J. Smith, 
M. Foss-Feig, S. Michalakis, A. V. Gorshkov, and 
C. Monroe, Nature {\bf 511}, 198 (2014).

\bibitem{exp2}
P. Jurcevic, B. P. Lanyon, P. Hauke, C. Hempel, P. Zoller, R. Blatt, 
and C. F. Roos, Nature {\bf 511}, 202 (2014).

\bibitem{tech5}
J. S. Douglas, H. Habibian, A. V. Gorshkov, H. J. Kimble, and D. E. Chang, 
Nature Photon. {\bf 9}, 326 (2015).

\bibitem{Saffman2010}
M. Saffman, T. G. Walker, and K. M\o lmer, Rev. Mod. Phys. {\bf 82}, 2313 (2010).

\bibitem{tech2} C. Schneider, D. Porras, and T. Schaetz, Rep. Prog. Phys.
{\bf 75}, (2012).

\bibitem{tech3} A. Bermudez, T. Schaetz, and M. B. Plenio, Phys. Rev. Lett.
{\bf 110}, 110502 (2013).

\bibitem{exp0}
E. Shahmoon and G. Kurizki, Phys. Rev. A {\bf 87}, 033831 (2013).
 
\bibitem{tech4}
T. Grass and M. Lewenstein, Europhys. Journ. Quantum Techn. {\bf 1}, 
8 (2014).

\bibitem{maghrebi2015tris}
Z.-X. Gong, M. F. Maghrebi, A. Hu, M. Foss-Feig, P. Richerme, C. Monroe, and A. V. Gorshkov,
Phys. Rev. B {\bf 93}, 205115 (2016).

\bibitem{dif}
P. Di Francesco, P. Mathieu, and D. Senechal, {\em Conformal
Field Theory} (New York, Springer, 1997). 

\bibitem{muss}
G. Mussardo, {\em Statistical Field Theory: 
An Introduction to Exactly Solved Models in Statistical Physics} 
(Oxford University Press, 2010).

\bibitem{polyakov_1970}
A. M. Polyakov, JETP Lett. {\bf 12}, 381 (1970).

\bibitem{cardy_1994}
J. Cardy, J. Phys. A: Math. Gen. {\bf 25}, L201 (1994).

\bibitem{langlands_1994}
R. Langlands, P. Pouliot, and Y. Saint-Aubin, 
Bull. Am. Math. Soc. {\bf 30}, 1 (1994).

\bibitem{deng_2002}
Y. Deng Y and H. W. J. Bl\"ote, Phys. Rev. Lett. {\bf 88}, 190602 (2002).

\bibitem{gori_2015}
G. Gori and A. Trombettoni, J. Stat. Mech. P07014 (2015).

\bibitem{penedones_2015}
C. Cosme, J. M. Viana Parente Lopes, and J. Penedones, JHEP \textbf{8}, 22 (2015).  

\bibitem{sak_1973}
J. Sak, Phys. Rev. B {\bf 8}, 1 (1973). 

\bibitem{spohn1999}
H. Spohn and W. Zwerger, J. Stat. Phys. {\bf 94} 5 (1999).

\bibitem{MW}
N. D. Mermin and H. Wagner, Phys. Rev. Lett. {\bf 17} 1133 (1966).

\bibitem{lebellac}
M. Le Bellac, {\em Quantum and statistical field theory} (Clarendon Press, 1992).

\bibitem{defenu2014}
N. Defenu, P. Mati, I. G. Marian, I. Nandori, and A. Trombettoni,
JHEP 1505(141) (2014).
 
\bibitem{dyson1969}
 F. J. Dyson, Comm. Math. Phys. 12, {\bf 91} (1969).
 
\bibitem{thouless1969}
  D. J. Thouless, Phys. Rev. 187, {\bf 732} (1969).
  
\bibitem{anderson1970}
 P. W. Anderson, G. Yuval, and D. R. Hamann, Phys. Rev. B {\bf 1},
4464 (1970).

\bibitem{cardy1981}
 J. L. Cardy, J. Phys. A {\bf 14}, 1407 (1981).

\bibitem{frolich1982}
J. Frolich and T. Spencer, Comm. Math. Phys. {\bf 84}, 87 (1982). 

\bibitem{lui2001}
E. Luijten and H. Messingfeld, Phys. Rev. Lett. {\bf 86}, 5305
(2001).

\bibitem{lui1997} E. Luijten, Ph.D. thesis, Delft University of Technology (1997).


\bibitem{Pientka2013}
F. Pientka, L. I. Glazman, and F. von Oppen, Phys. Rev.  B {\bf 88}, 155420 (2013).

\bibitem{Pientka2014}
F. Pientka, L. I. Glazman, and F. von Oppen, Phys. Rev. B {\bf 89}, 180505 (2014).

\bibitem{kitaev}
A. Y. Kitaev, Phys. Usp. {\bf 44}, 131 (2001).

\bibitem{libro_cha} 
A. Dutta, G. Aeppli, B. K. Chakrabarti, U. Divakaran, T. F. Rosenbaum, 
and D. Sen, {\em Quantum Phase Transitions in Transverse Field Models} 
(Cambridge University Press, 2015).

\bibitem{laflorencie}
N. Laflorencie, I. Affleck, and M. Berciu,
J. Stat. Mech. P12001 (2005).

\bibitem{defenu}
N. Defenu, A. Trombettoni, and A. Codello, Phys. Rev. E {\bf 92}, 052113 (2015). 

\bibitem{Maghrebi2015}
M. F. Maghrebi, Z.-X. Gong, M. Foss-Feig, and A. V. Gorshkov, 
Phys. Rev. B {\bf 93}, 125128 (2016).

\bibitem{huang}
K. Huang, {\em Statistical Mechanics}, 2nd  ed. (New York, Wiley, 1987).

\bibitem{wilczek}
C. Holzhey, F. Larsen, and F. Wilczek, Nucl. Phys. B {\bf 424}, 443 (1994).
  
\bibitem{calabrese}
P. Calabrese and J. Cardy, J. Stat. Mech. P06002 (2004).

\bibitem{ares}
F. Ares, J. G. Esteve, F. Falceto, and A. R. de Queiroz, Phys. Rev. A {\bf 92}, 042334 (2015); 
J. Stat. Mech. 043106 (2016).

\bibitem{grad}
I. S. Gradshteyn and I. M. Ryzhik, {\em 
Tables of Integrals, Series, and Products} (New  York, Academic, 2007).
 
\bibitem{abr}
M. Abramowitz and I. A. Stegun, {\em Handbook of Mathematical Functions} 
(New York, Dover, 1964). 
 
\bibitem{nist}
F. W. J. Olver, D. W. Lozier, R. F. Boisvert, and C. W. Clark, 
{\em NIST Handbook of Mathematical Functions} (Cambridge University Press, 2010).

\bibitem{continentino}
M. A. Continentino, F. Deus, and H. Caldas,
Phys. Lett. A {\bf 378}, 1561 (2014).

\bibitem{shankar}
R. Shankar, Rev. Mod. Phys. {\bf 66}, 129 (1994).

\bibitem{janos}
J. Polonyi, Centr. Eur. J. Phys. {\bf 1}, 1 (2003).

\bibitem{germandis}
A. Rodriguez, V. A. Malyshev, G. Sierra, M. A. Martin-Delgado, J. Rodriguez-Laguna, and F. Dominguez-Adame,
Phys. Rev. Lett. {\bf 90}, 027404 (2003).

\bibitem{majorana} 
E. Majorana, Nuovo Cimento {\bf 5}, 171 (1937).

\bibitem{pal}
P. B. Pal, Am. J. Phys. {\bf 79}, 485 (2011).

\bibitem{peskin}
M. E. Peskin and D. V. Schroeder, 
{\em An Introduction To Quantum Field Theory} (Reading, Addison-Wesley, 1995). 

\bibitem{wei1}
S. Weinberg, {\em The Quantum Theory of Fields}, Vol. 1 
(Cambridge University Press, 1995).

\bibitem{Magh_bis}
M. F. Maghrebi, Z.-X. Gong, and A. V. Gorshkov, \verb|arXiv:1510.01325|

\bibitem{Ablowitz2003}
M. J. Ablowitz and A. T. S. Fokas, {\em Complex Variables Introduction and 
Applications} (Cambridge University Press, 2003).

\bibitem{dutta2001}
A. Dutta and J. K. Bhattacharjee, Phys. Rev. B {\bf 64},
184106 (2001).

\bibitem{cardybook}
J. Cardy, {\em Scaling and Renormalization in Statistical Physics}, Cambridge Lecture Notes in Physics (1996).

\bibitem{fred}
K. Fredenhagen, Comm. Math. Phys. {\bf 97}, 461 (1985).

\bibitem{nota2}
Another general way to achieve the same conclusion is to consider a  spectral expansion in form factors (FFE), valid no matter the dimensionality of the system and the strength of the interaction between the excitations \cite{muss}.
It is easy to see that in our case the validity of FFE is invalidated by the loss of invariance under Euclidean rotations.

\bibitem{gori_15}
G. Gori, S. Paganelli, A. Sharma, P. Sodano, and A. Trombettoni, 
Phys. Rev. B {\bf 91}, 245138 (2015). 

\bibitem{eisert}
J. Eisert, M. Cramer, and M. B. Plenio, Rev. Mod. Phys. {\bf 82}, 277 (2010).

\bibitem{callan}
C. Callan and F. Wilczek, Phys. Lett. B {\bf 333}, 55 (1994).

\bibitem{casini}
H. Casini and M. Huerta, J. Phys. A: Math. Gen. {\bf 42}, 504007 (2009).

\bibitem{peschel}
I. Peschel, \verb|arXiv:1109.0159|, Braz. J. Phys. {\bf 42}, 267 (2012).

\bibitem{nez}
M. Ghasemi Nezhadhaghighi and M. A. Rajabpour,
Europhys. Lett. {\bf 100}, 60011 (2012).

\bibitem{refpar}
T. D. Schultz, D. C. Matthis, and E. H. Lieb,
Rev. Mod. Phys. {\bf 36}, 856 (1964).

\bibitem{delfino04}
G. Delfino, J. Phys. A {\bf 37} (2004), R 45.

\bibitem{um2006}
J. Um, S.-I. Lee, and B. J. Kim, 
J. Kor. Phys. Soc. {\bf 50}, 1 (2007).

\bibitem{slava}
M. F. Paulos, S. Rychkov, B. C. van Rees, and B. Zan,
Nucl. Phys. B {\bf 902}, 246 (2016).

\bibitem{wouters2015}
M. Van Regemortel, D. Sels, and M. Wouters, 
Phys. Rev. A {\bf 93}, 032311 (2016).

\bibitem{sak}
J. J. Sakurai, {\em Advanced Quantum Mechanics} (Reading, Addison Wesley, 1967).

\bibitem{birrell}
N. D. Birrell and  P. C. W. Davies, {\em Quantum fields in curved space} (Cambridge University Press, 1982).

\bibitem{landau2}
L. D. Landau and E. M. Lifshitz, {\em The Classical Theory of Fields} (Oxford, Pergamon Press, 1971). 


\end{thebibliography}
\end{document}